\documentclass[twocolumn,amsmath,amssymb,floatfix,nature,shownopacs,footinbib,superscriptaddress]{revtex4-1}
\usepackage{amsmath,amssymb,natbib,bm,graphicx,url,epsf,color}
\usepackage[british]{babel}
\usepackage{wasysym}
\usepackage{MnSymbol}
\setcitestyle{super} 

\begin{document}

\title{Controlling charge quantization with quantum fluctuations}

\author{S. Jezouin}
\email{These authors contributed equally to this work.}
\affiliation{Centre de Nanosciences et de Nanotechnologies (C2N), CNRS, Univ Paris Sud-Universit\'e Paris-Saclay, Univ Paris Diderot-Sorbonne Paris Cit\'e, 91120 Palaiseau, France}
\author{Z. Iftikhar}
\email{These authors contributed equally to this work.}
\affiliation{Centre de Nanosciences et de Nanotechnologies (C2N), CNRS, Univ Paris Sud-Universit\'e Paris-Saclay, Univ Paris Diderot-Sorbonne Paris Cit\'e, 91120 Palaiseau, France}
\author{A. Anthore}
\affiliation{Centre de Nanosciences et de Nanotechnologies (C2N), CNRS, Univ Paris Sud-Universit\'e Paris-Saclay, Univ Paris Diderot-Sorbonne Paris Cit\'e, 91120 Palaiseau, France}
\author{F.D. Parmentier}
\affiliation{Centre de Nanosciences et de Nanotechnologies (C2N), CNRS, Univ Paris Sud-Universit\'e Paris-Saclay, Univ Paris Diderot-Sorbonne Paris Cit\'e, 91120 Palaiseau, France}
\author{U. Gennser}
\affiliation{Centre de Nanosciences et de Nanotechnologies (C2N), CNRS, Univ Paris Sud-Universit\'e Paris-Saclay, Univ Paris Diderot-Sorbonne Paris Cit\'e, 91120 Palaiseau, France}
\author{A. Cavanna}
\affiliation{Centre de Nanosciences et de Nanotechnologies (C2N), CNRS, Univ Paris Sud-Universit\'e Paris-Saclay, Univ Paris Diderot-Sorbonne Paris Cit\'e, 91120 Palaiseau, France}
\author{A. Ouerghi}
\affiliation{Centre de Nanosciences et de Nanotechnologies (C2N), CNRS, Univ Paris Sud-Universit\'e Paris-Saclay, Univ Paris Diderot-Sorbonne Paris Cit\'e, 91120 Palaiseau, France}
\author{I.P. Levkivskyi}
\affiliation{Institute for Theoretical Physics, ETH Zurich, CH-8093 Zurich, Switzerland}
\author{E. Idrisov}
\affiliation{D\'epartement de Physique Th\'eorique, Universit\'e de Gen\`eve, CH-1211 Gen\`eve, Switzerland}
\author{E.V. Sukhorukov}
\affiliation{D\'epartement de Physique Th\'eorique, Universit\'e de Gen\`eve, CH-1211 Gen\`eve, Switzerland}
\author{L.I. Glazman}
\affiliation{Department of Physics, Yale University, New Haven, CT 06520, USA}
\author{F. Pierre\thanks{frederic.pierre@lpn.cnrs.fr}}
\email[e-mail: ]{frederic.pierre@u-psud.fr}
\affiliation{Centre de Nanosciences et de Nanotechnologies (C2N), CNRS, Univ Paris Sud-Universit\'e Paris-Saclay, Univ Paris Diderot-Sorbonne Paris Cit\'e, 91120 Palaiseau, France}

\maketitle

{\sffamily
In 1909, Millikan showed that the charge of electrically isolated systems is quantized in units of the elementary electron charge $e$. 
Today, the persistence of charge quantization in small, weakly connected conductors allows for circuits where single electrons are manipulated, with applications in e.g. metrology, detectors and thermometry \cite{Schon1990,Ingold1992,Likharev1999,Meschke2011,Pekola2013}. 
However, quantum fluctuations progressively reduce the discreteness of charge as the connection strength is increased. 
Here we report on the full quantum control and characterization of charge quantization.
By using semiconductor-based tunable elemental conduction channels to connect a micrometer-scale metallic island, the complete evolution is explored while scanning the entire range of connection strengths, from tunnel barrier to ballistic contact. 
We observe a robust scaling of charge quantization as the square root of the residual electron reflection probability across a quantum channel when approaching the ballistic critical point, which also applies beyond the regimes yet accessible to theory \cite{Flensberg1993,Matveev1995,Nazarov1999}.
At increased temperatures, the thermal fluctuations result in an exponential suppression of charge quantization as well as in a universal square root scaling, for arbitrary connection strengths, in agreement with expectations \cite{Nazarov1999}.
Besides direct applications to improve single-electron functionalities and for the metal-semiconductor hybrids emerging in the quest toward topological quantum computing \cite{Albrecht2016}, the knowledge of the quantum laws of electricity will be essential for the quantum engineering of future nanoelectronic devices.
}

Some of the most fundamental theoretical predictions have so far eluded experimental confirmation.
Charging effects are generally found to diminish as the contacts' conductances are increased \cite{Kouwenhoven1991,Staring1991,vanderVaart1993,Molenkamp1995,Joyez1997,Chouvaev1999,Berman1999,Duncan1999,Amasha2011}.
However, while some measurements support the fundamental prediction \cite{Flensberg1993,Matveev1995,Nazarov1999} that charge quantization vanishes in the presence of one ballistic channel \cite{Kouwenhoven1991,Staring1991,vanderVaart1993,Duncan1999}, others conclude the opposite \cite{Pasquier1993,Crouch1996,Liang1998,Cronenwett1998,Tkachenko2001,Amasha2011}.
Unsurprisingly, the scaling behavior predicted for the reduction of charge quantization \cite{Flensberg1993,Matveev1995,Nazarov1999} has also remained, up to now, elusive despite several attempts \cite{Berman1999,Duncan1999}.

\begin{figure}[!tbh]
\centering\includegraphics[width=\columnwidth]{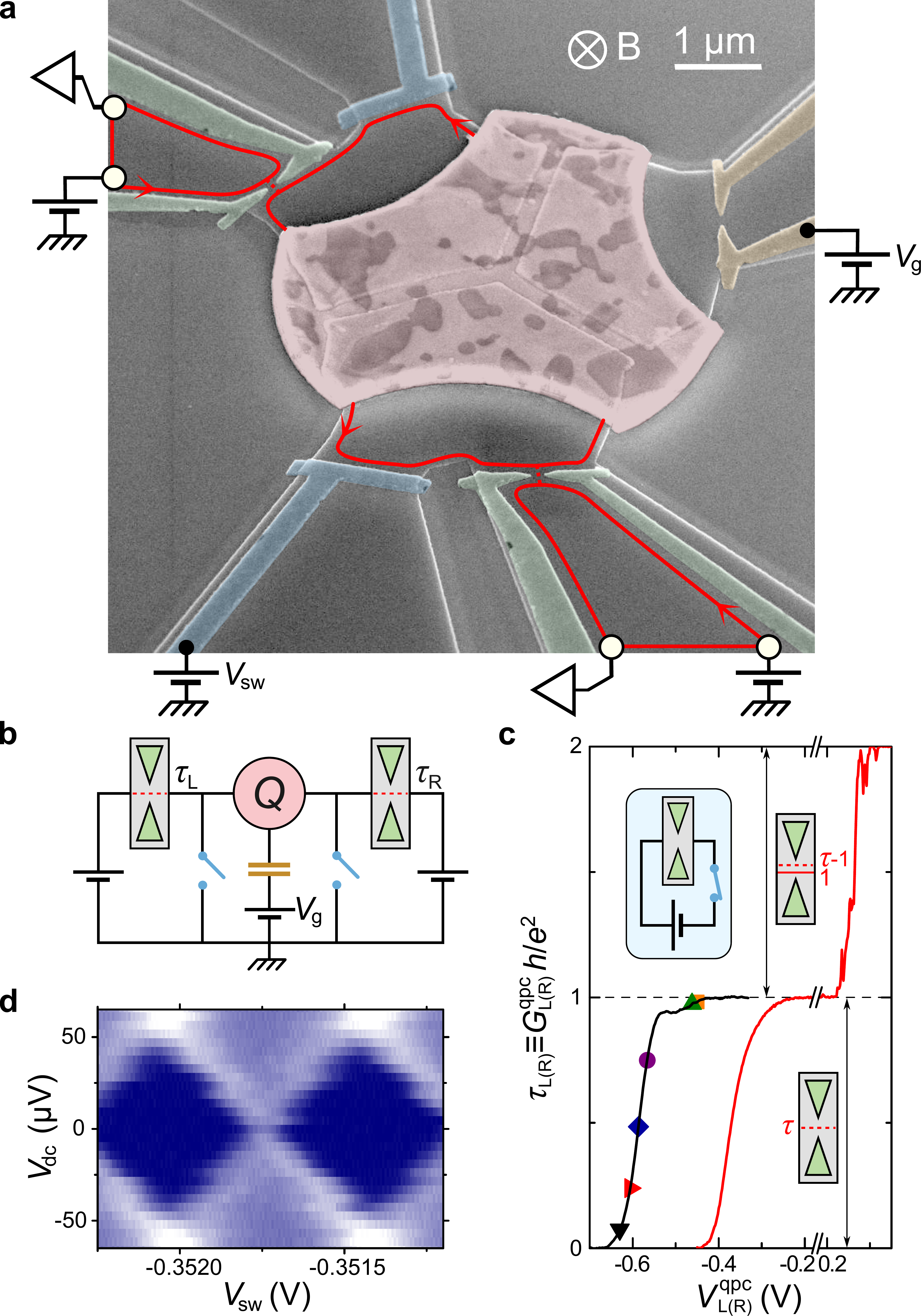}
\caption{
\footnotesize
\textbf{Tunable quantum connection to a metallic island}.  
\textbf{a}, Colored sample micrograph. 
A micrometer-scale metallic island (red) is connected to large electrodes (white circles) through two quantum point contacts (QPCs, green split gates) formed in a buried 2D electron gas (2DEG, darker gray delimited by bright lines).
The lateral gates (blue) implement short-circuit switches as shown in (b).
The top-right yellow gates, tuned to deplete the 2DEG underneath, are capacitively coupled to the island.
In the applied $B\simeq4~$T, the current propagates along two edge channels (red line) in the direction indicated by arrows.
\textbf{b}, Sample schematic. 
\textbf{c}, The `intrinsic' (i.e. switch closed) conductance $G^\mathrm{qpc}_\mathrm{L(R)}$ across the top-left QPC$_L$ (bottom-right QPC$_R$) is shown versus split gate voltage $V^\mathrm{qpc}_\mathrm{L(R)}$ as a black (red) line.
Symbols indicate the set-points of QPC$_L$ used thereafter.
\textbf{d}, Coulomb diamond patterns in the device conductance $G_\mathrm{SET}$ (larger shown brighter, from 0 in dark blue up to 0.13$e^2/h$ in white) measured versus gate ($V_{\mathrm{sw}}$) and bias ($V_\mathrm{dc}$) voltages for tunnel contacts ($\tau_{L,R}\ll 1$). 
}
\label{fig1}
\end{figure}

A plausible explanation of the varying results regarding the charge quantization criteria is that, in the previously investigated devices, the quantum channels and the conductor were not completely distinct circuit elements.
With a small island, in which the density of states is discrete, the non-local electronic wave functions merge the connected channels and the island into a complex quantum conductor, where Coulomb interactions may play a non-trivial role.
As a result, charging effects can develop even if one of the conduction channels taken separately is perfectly ballistic. 
This phenomenon is called mesoscopic Coulomb blockade \cite{Aleiner1998,Cronenwett1998,Amasha2011}.

Investigating charge quantization at the most elemental single-channel level therefore requires tunable conduction channels linked to a conductor with a negligible electronic level spacing.
Although this can be realized by increasing the island's size, the latter must remain small enough to preserve charge quantization. 
Indeed, thermal fluctuations average out charge quantization unless the charging energy associated with the addition of one electron in the island, $E_C=e^2/2C$, where the island's geometrical capacitance $C$ increases with size, is larger than the thermal energy $k_BT$, with $k_B$ the Boltzmann constant and $T$ the temperature \cite{Schon1990,Ingold1992}.

\begin{figure}[!tbh]
\centering\includegraphics[width=1\columnwidth]{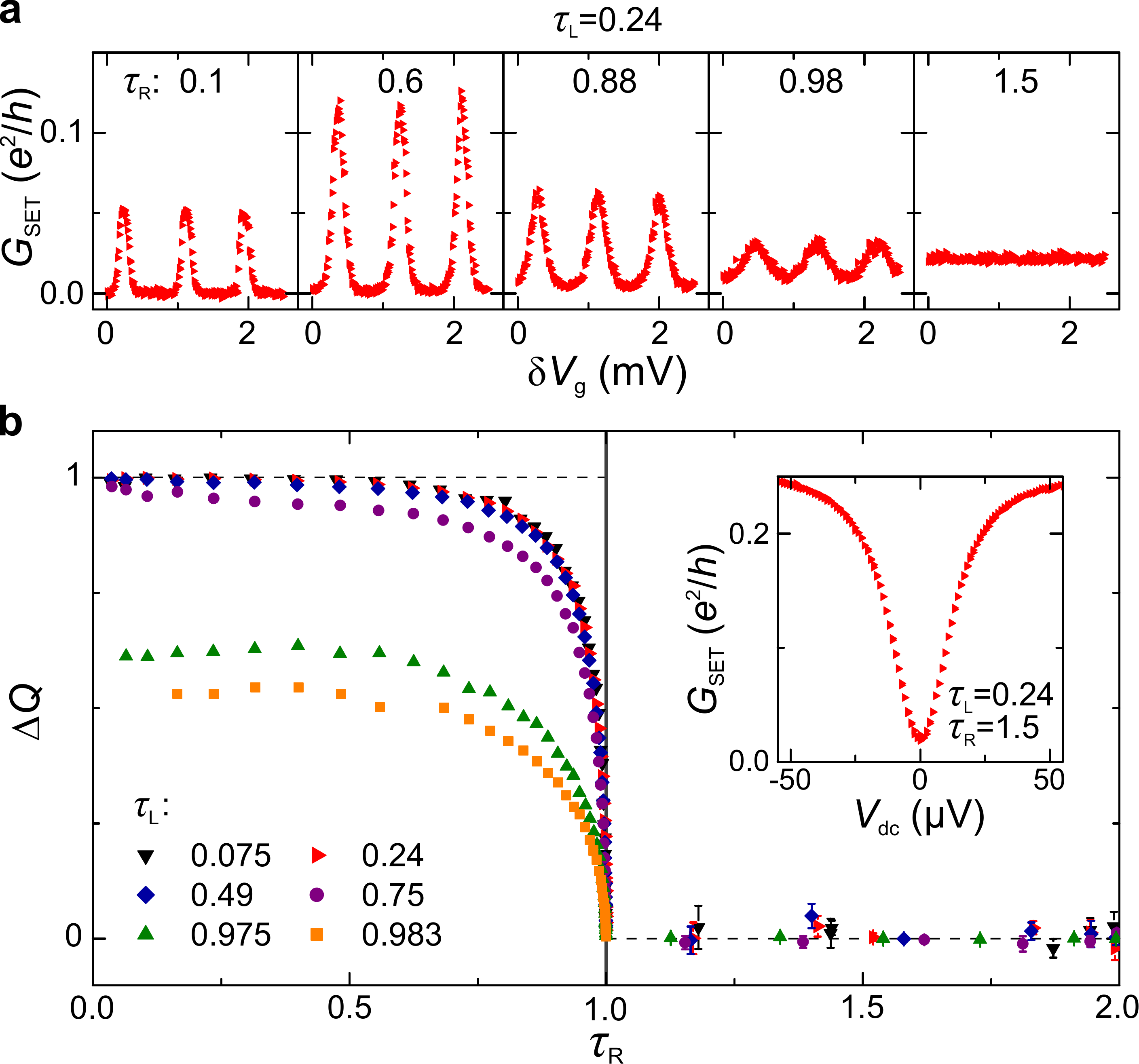}
\caption{
\footnotesize
\textbf{Charge quantization versus connection strength} at $T\simeq17~$mK. 
\textbf{a}, Conductance sweeps $G_\mathrm{SET}(\delta V_\mathrm{g})$ with a fixed $\tau_L=0.24$, and $\tau_R=0.1$, $0.6$, $0.88$, $0.98$ and $1.5$, from left to right respectively.
\textbf{b}, Visibility of $G_\mathrm{SET}$ oscillations ($\Delta Q\equiv(G_\mathrm{SET}^\mathrm{max}-G_\mathrm{SET}^\mathrm{min})/(G_\mathrm{SET}^\mathrm{max}+G_\mathrm{SET}^\mathrm{min})$) versus $\tau_R$, with each set of symbols corresponding to a different QPC$_L$ set-point.
\textbf{Inset}, Dynamical Coulomb blockade renormalization of $G_\mathrm{SET}$ versus dc voltage in the absence of charge quantization, at $\tau_L=0.24$ and $\tau_R=1.5$.
}
\label{fig2}
\end{figure}

We have solved these conflicting requirements with the hybrid metal-semiconductor single-electron transistor (SET) shown in Fig.~1a, implementing the schematic circuit of Fig.~1b:
A central metallic island with a continuous density of states (colored red) is connected to large electrodes (represented by white disks) through two Ga(Al)As quantum point contacts (QPC$_{L,R}$) that emulate single-channel quantum conductors over the entire range of coupling strengths.

The metallic island, made of a metallic AuGeNi alloy, has a negligible electronic level spacing $\delta\approx k_B\times0.2~\mu$K, five orders of magnitude smaller than the base electronic temperature $T\simeq17~$mK.
It is galvanically connected, by thermal annealing, to a 105~nm deep Ga(Al)As high mobility two-dimensional electron gas (2DEG, darker grey areas delimited by bright lines in Fig.~1a).
Achieving an almost perfectly transparent metal-2DEG electrical contact is crucial to reach the ballistic channel limit.
Remarkably, the reflection probability of electrons at the interface is here below $0.05\%$.

The QPCs are located in the 2DEG and tuned by field effect with the voltage applied to capacitively coupled metallic split gates (colored green; the top-right split gates colored yellow are negatively biased to remove the 2DEG underneath).
Besides tuning, the precise characterization of each QPC, independently, is necessary for the quantitative exploration of charge quantization versus connection strength.
However, in the SET configuration the QPC conductances are interconnected and renormalized by Coulomb blockade.
Moreover, only their series combination is accessible.
In order to completely characterize QPC$_{L,R}$, we have implemented with adjacent gates (colored blue) the on-chip switches shown in Fig.~1b. 
The measured $\tau_{L,R}\equiv G^\mathrm{qpc}_{L,R}h/e^2$, with $h$ the Planck constant and $G^\mathrm{qpc}_{L,R}$ the conductances of QPC$_{L,R}$ switches closed (inset of Fig.~1c), directly give the `intrinsic' (not renormalized by Coulomb blockade) transmission probabilities of the constitutive quantum channels, which fully characterize the connection strength to the metallic island.
As illustrated in Fig.~1c, $\tau_{L(R)}\leq1$ corresponds to a single (spin-polarized, see below) channel of transmission probability $\tau_{L(R)}$ across QPC$_{L(R)}$.
For $1<\tau_R\leq2$, there are two channels across QPC$_R$, one fully ballistic and the other one of transmission probability $\tau_R-1$.
With this approach, we achieve a remarkable accuracy, down to $0.1\%$ near the ballistic limit.

The sample is immersed into a perpendicular magnetic field $B\simeq4~$T corresponding to the integer quantum Hall effect at filling factor $\nu=2$.
In this regime, the electrical current propagates along two edge channels (shown as a single red line in Fig.~1a) in the direction indicated by arrows, which does not influence charge quantization (for a specific discussion see Methods, section `Conductance in the near ballistic regime with strong thermal fluctuations', part A).
The large Zeeman splitting results in the full separation between the successive openings of the two spin-polarized quantum channels across the QPCs (Fig.~1c).

Charge quantization in the central island is unequivocally evidenced from periodic oscillations of the SET differential conductance $G_\mathrm{SET}$ (across QPC$_L$-island-QPC$_R$) when sweeping a capacitively coupled gate voltage, which develop into Coulomb diamonds with dc bias voltage $V_\mathrm{dc}$ (Fig.~1d).
With both QPCs in the tunnel regime, $\tau_{L,R}\ll1$, the diamonds' extension in $V_\mathrm{dc}$ gives the charging energy $E_C\simeq k_B\times0.3~$K ($C\simeq3.1~$fF).  

We first probe the evolution of charge quantization with transmission probability directly from $G_\mathrm{SET}$ raw periodic modulations.
Figure 2a displays as symbols $G_\mathrm{SET}$ measured at $T\simeq17~$mK and $V_\mathrm{dc}=0$ while sweeping the capacitively coupled $V_\mathrm{g}$, for QPC$_L$ fixed to $\tau_L=0.24$ and with each panel corresponding to a different QPC$_R$ tuning ($\tau_R=0.1$, $0.6$, $0.88$, $0.98$ and $1.5$, from left to right respectively).
These raw data reveal the remarkable robustness of charge quantization to connection strength.
At $\tau_R=0.1$ and $0.6$, the presence of sharp periodic peaks separated by $G_\mathrm{SET}\approx0$ intervals signals an essentially unaltered charge quantization over the greater part of transmission probabilities.
While $G_\mathrm{SET}(\delta V_\mathrm{g})$ progressively evolves with increasing $\tau_R<1$ into a sinusoid with non-zero minima, relatively important modulations of fixed ($\tau_R$ independent) period persist very close to the ballistic limit, at $\tau_R=0.98$.
In stark contrast, $G_\mathrm{SET}$ is independent of $V_\mathrm{g}$ at $\tau_R=1.5$, confirming the predicted complete collapse of charge quantization in the presence of a fully ballistic channel.
Note that $G_\mathrm{SET}$ remains reduced by Coulomb interactions, even at $\tau_R=1.5$, as evidenced from the pronounced conductance dip at low $V_\mathrm{dc}$ (inset of Fig.~2b).
Indeed, the so-called dynamical Coulomb blockade does not rely on a quantized island charge, but results from the discreteness of charge transfers across non-ballistic channels \cite{Schon1990,Ingold1992}.

The degree of charge quantization versus connection strength is characterized, separately from the channels' dynamical Coulomb blockade renormalization, by focusing on the periodic modulations' visibility $\Delta Q\equiv (G_\mathrm{SET}^\mathrm{max}-G_\mathrm{SET}^\mathrm{min})/(G_\mathrm{SET}^\mathrm{max}+G_\mathrm{SET}^\mathrm{min})$, with $G_\mathrm{SET}^\mathrm{max(min)}$ the maximum (minimum) SET conductance over one gate voltage period and, from now on, $V_\mathrm{dc}=0$.
A visibility $\Delta Q=1(0)$ clearly signals a full (an absence of) charge quantization.
Moreover, the visibility $\Delta Q$ is directly proportional to the island's charge oscillations with gate voltage (i.e. charge quantization) when one channel approaches the ballistic limit (e.g. $\tau_R\rightarrow1$) \cite{Matveev1995,Furusaki1995b,Yi1996,LeHur2002}.
As put forward in Ref.~\citenum{Yi1996}, this proportionality coefficient reduces to the numerical factor $e/(2\pi 1.59)$ for $\tau_L\ll1$ and $k_BT\ll E_C$.

Figure~2b shows $\Delta Q$ versus $\tau_R$ at $T\simeq17$~mK, with each set of symbols corresponding to a different tuning of the second QPC ($\tau_L\in\{0.075,$ $0.24,$ $0.49,$ $0.75,$ $0.975,$ $0.983\})$. 
The robustness of charge quantization with the connection strength of one channel ($\tau_R$) is established now independently of the second channel ($\tau_L$), from the nearly constant $\Delta Q$ for $\tau_R\lesssim0.6$.
When further increasing $\tau_R$, $\Delta Q$ noticeably diminishes and systematically collapses to zero precisely at the ballistic critical point $\tau_R=1$.
At $\tau_R\geq1$, in the presence of one ballistic channel, $\Delta Q$ remains perfectly null at experimental accuracy (see Methods for additional tests).

\begin{figure}[!htb]
\centering\includegraphics[width=1\columnwidth]{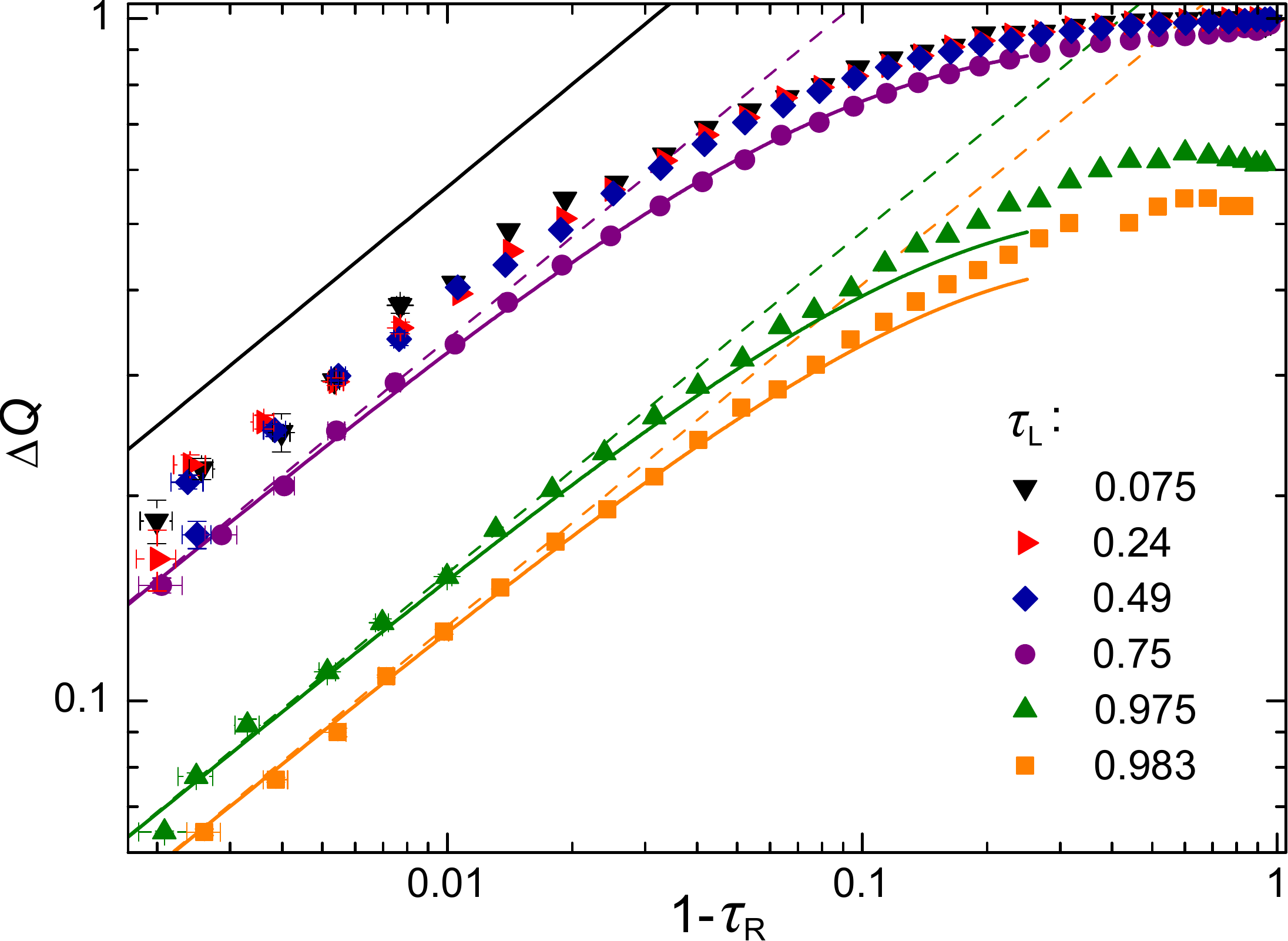}
\caption{
\footnotesize
\textbf{Charge quantization scaling near the ballistic critical point}. 
The $\Delta Q$ data at $T\simeq17$~mK are displayed versus $1-\tau_R$ in a $\log$-$\log$ scale, with distinct sets of symbols for the different QPC$_L$ set-points. 
Continuous lines are quantitative predictions (no fit parameters) derived assuming $k_BT\ll E_C$, $1-\tau_R\ll1$, and either $\tau_L\ll1$ (top continuous line) or $1-\tau_L\ll1$ (three bottom continuous lines).
The power law $\Delta Q\propto\sqrt{1-\tau_R}$ (straight lines) is systematically observed at $1-\tau_R\lesssim0.02$, also at intermediate $\tau_L$.
}
\label{fig3}
\end{figure}

Power laws characterizing the scaling of charge quantization as $\tau_R\rightarrow1$ are best revealed by plotting in a $\log$-$\log$ scale $\Delta Q$ versus the `distance' $1-\tau_R>0$ from the ballistic critical point.
As shown in Fig.~3, the $T=17$~mK data (symbols) systematically vanish as $\sqrt{1-\tau_R}$ (straight lines) for $1-\tau_R\lesssim0.02$.

\begin{figure*}[!thb]
\centering\includegraphics[width=2\columnwidth]{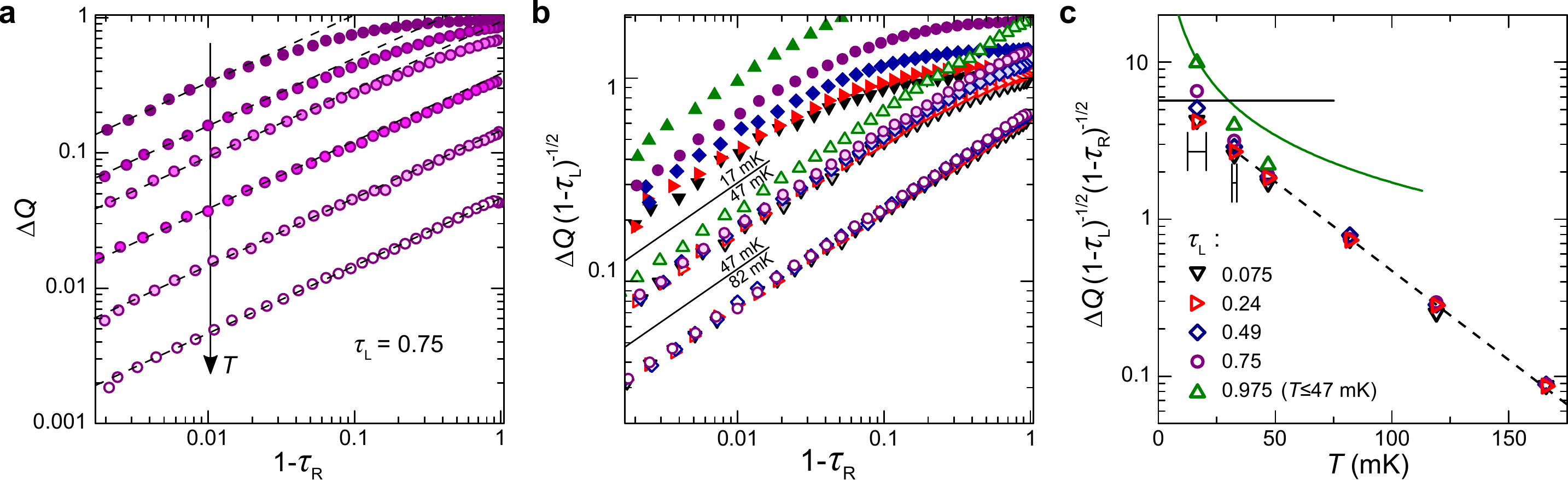}
\caption{
\footnotesize
\textbf{Crossover to a universal charge quantization scaling} as temperature is increased. 
\textbf{a}, Symbols display $\Delta Q$ versus $1-\tau_R$ at $\tau_L=0.75$ and for $T\simeq17$, 32, 47, 82, 119 and 166~mK, from top to bottom respectively.
The $\tau_R$ range over which $\Delta Q\propto\sqrt{1-\tau_R}$ (straight lines) extends up to the full interval $\tau_R\in[0,1]$ when increasing $T$.
\textbf{b}, The rescaled $\Delta Q/\sqrt{1-\tau_L}$ is shown versus $1-\tau_R$, with distinct set of symbols corresponding to different QPC$_L$ set-points as in (c).
Continuous lines separate the data at $T\simeq17$ (top, darker filling), 47 (middle) and 82~mK (bottom, brighter filling).
At $T=82$~mK, all the data collapse on a single universal curve $\Delta Q\propto\sqrt{(1-\tau_L)(1-\tau_R)}$.
\textbf{c}, Symbols display versus $T$, in semi-$\log$ scale, the fully rescaled data $\Delta Q/\sqrt{(1-\tau_L)(1-\tau_R)}$, extracted in the regime of small enough $1-\tau_R$ such that $\Delta Q\propto\sqrt{1-\tau_R}$.
Horizontal error bars represent the experimental temperature uncertainty at $T=17\pm4$~mK and $32\pm1$~mK.
Continuous lines are the quantitative predictions in the quantum regime $k_BT\ll E_C$, given by Eq.~1 (black) and Eq.~2 (green).
The straight dashed line displays an exponential decay close to predictions in the presence of strong thermal fluctuations (see text).
}
\label{fig4}
\end{figure*}

The Coulomb blockade theory of electronic transport in the presence of a nearly ballistic channel ($1-\tau_R\ll1$) relies on the bosonization approach initially developed to address correlated electrons at 1D.
Quantitative predictions were obtained for $k_BT\ll E_C$ and for a second channel either in the tunnel ($\tau_L\ll1$) or almost ballistic ($1-\tau_L\ll1$) regime \cite{Furusaki1995b,Matveev2002}.  
In both cases, $\Delta Q$ is expected to vanish as $\sqrt{1-\tau_R}$:
\begin{align}
\Delta&Q(1-\tau_R\ll1;\tau_L\ll1,k_BT\ll E_C)\simeq5.7\sqrt{1-\tau_R},\\
\Delta&Q(\sqrt{1-\tau_{L,R}}\ll\frac{k_BT}{E_C}\ll1)\simeq\frac{0.57E_C}{k_BT}\sqrt{(1-\tau_L)(1-\tau_R)}.
\end{align}
Note that such a scaling, initially proposed in Ref.~\citenum{Flensberg1993}, was also predicted for the gate voltage modulation of thermodynamic quantities for multi-channel junctions using an extension \cite{Nazarov1999} of the instanton technique \cite{Korshunov1987,Schon1990}. 

Remarkably, the data establishes the $\sqrt{1-\tau_R}$ scaling for arbitrary $\tau_L\in[0,1]$, beyond the tunnel and ballistic limits yet accessible to transport theory.
The dashed lines in Fig.~3 display the asymptotic ($\sqrt{1-\tau_{L,R}}\ll k_BT/E_C$) quantitative predictions of Eq.~2 at $T=17$~mK, without fitting parameters for our completely characterized device.
The non-asymptotic $\Delta Q$ predictions (Eq.~1 for $\tau_L\ll1$, see Methods for $1-\tau_L\ll1$) are shown as continuous lines versus $1-\tau_R<0.25$. 
Data and quantitative predictions are indistinguishable at $1-\tau_R\lesssim0.1$ for $\tau_L=0.983$, $0.975$ and also, more surprisingly, $0.75$.
Note that Eq.~1 prediction (black line) remains noticeably ($\sim25\%$) above the $\tau_L=0.075$ data at $1-\tau_R\ll1$.
This numerical difference could result from the finite experimental $T$, since Eq.~1 is exact only at $T=0$.

How does the combination of thermal and quantum fluctuations impact the quantization of charge?
As temperature rises, the population of additional charge states is expected to average out charge quantization \cite{Schon1990,Ingold1992}.
Figure~4a displays as symbols $\Delta Q$ measured versus $1-\tau_R$ at different temperatures, from $T=17$~mK (darker filling) to $166$~mK (brighter filling), for the representative QPC$_L$ setting $\tau_L=0.75$.
As naively expected, $\Delta Q$ diminishes as $T$ increases.
Remarkably, in line with thermodynamic expectations \cite{Nazarov1999} (Methods), the $\Delta Q\propto\sqrt{1-\tau_R}$ scaling (straight lines) that originates from quantum fluctutations not only persists for increasing $T$, but extends over a widening range of $\tau_R$ up to the full scale $\tau_R\in[0,1]$.

The crossover toward this universal behaviour is established by comparing the rescaled visibility $\Delta Q/\sqrt{1-\tau_L}$ for different $\tau_L$ settings, versus $1-\tau_R$.
The symbols in Fig.~4b represent the rescaled data at $T=17$, 47 and 82~mK, with brighter fillings at higher temperatures.
As $T$ increases, the scatter associated with the various $\tau_L$ narrows.
Remarkably, for $T\geq82$~mK, the rescaled data collapse onto a single, universal, straight line $\Delta Q\propto\sqrt{(1-\tau_L)(1-\tau_R)}$ over the full range $\tau_{L,R}\in[0,1]$.

The temperature dependence is further characterized by plotting $\Delta Q/\sqrt{(1-\tau_L)(1-\tau_R)}$ (determined at low enough $1-\tau_R$ such that $\Delta Q\propto\sqrt{1-\tau_R}$) in semi-$\log$ scale versus temperature in Fig.~4c (symbols).
The $k_BT\ll E_C$ prediction of Eq.~1 (Eq.~2) is displayed as a black (green) continuous line for $T<75~(115)$~mK.
We find for $T\geq82$~mK (up to $166$~mK, $2.8\leq \pi^2k_BT/E_C\leq5.6$) that the different $\tau_L$ data points collapse on the same exponential decay $\Delta Q\sim\sqrt{(1-\tau_L)(1-\tau_R)}\exp(-0.8\pi^2k_BT/E_C)$ (dashed line).
We have extended the Coulomb blockade theory for the conductance to include thermal fluctuations in the limits of tunnel or nearly ballistic channels (Methods).
In the regime of strong thermal averaging, we predict $\Delta Q\propto\sqrt{(1-\tau_L)(1-\tau_R)}\exp(-\pi^2k_BT/E_C)$ (neglecting non-exponential $T$ factors), as also expected for thermodynamic properties \cite{Nazarov1999} (Methods), and in close agreement with the experimental findings regarding both the effect of $\tau_{L,R}$ and $T$.

Although theoretical predictions for low-temperature transport yet apply to the near ballistic and tunnel limits, we anticipate that recent advances, including those in numerical renormalization group \cite{Mitchell2016}, will open access to the full range of connection strengths.
Our results may therefore provide an auspicious test-bed for strongly correlated electron theoretical methods, for which these non-perturbative techniques are ubiquitous.
The understanding and on-demand control of charge quantization in mesoscopic circuits might lead to applications beyond the field of single-electronics.
Its central role in the different quantum laws of electricity with coherent conductors signals direct quantum engineering implications for future nanoelectronics.
These include semiconductor-metal hybrid devices, that emerge as crucial elements in the quest for topologically protected quantum bits \cite{Albrecht2016}.
The present hybrid implementation also opens the path to further fundamental explorations, including of charge quantization with correlated electrons, such as in the multi-channel Kondo regime and/or with fractionally charged anyonic quasiparticles.

\bibliographystyle{nature}

\vspace{\baselineskip}
\small
{\noindent\textbf{Acknowledgments.}}
This work was supported by the European Research Council (ERC-2010-StG-20091028, no. 259033), the French RENATECH network, the national French program `Investissements d'Avenir' (Labex NanoSaclay, ANR-10-LABX-0035), the U.S. Department of Energy (DE-FG02-08ER46482) and the Swiss National Science Foundation.

{\noindent\textbf{Author Contributions.}} 
S.J. and Z.I. performed the experiment with inputs from A.A. and F.P.;
S.J., Z.I., A.A. and F.P. analyzed the data;
F.D.P. fabricated the sample and contributed to a preliminary experiment;
U.G., A.C. and A.O. grew the 2DEG;
I.L., E.I., E.V. and L.G. developed the strong thermal fluctuations theory;
F.P. led the project and wrote the manuscript with inputs from all authors.

{\noindent\textbf{Author Information.}}
The authors declare no competing financial interests.
Correspondence and requests for materials should be addressed to F.P. (frederic.pierre@u-psud.fr).
\normalsize 

\newpage
{\Large\noindent\textbf{METHODS}}

\small

{\noindent\textbf{Sample.}} 
The sample is nanostructured by standard e-beam lithography in a GaAs/Ga(Al)As two-dimensional electron gas located 105~nm below the surface, of density $2.5\times10^{11}~\mathrm{cm}^{-2}$ and mobility $10^6~\mathrm{cm}^2\mathrm{V}^{-1}\mathrm{s}^{-1}$. 
The ohmic contact between micrometer-scale metallic island and buried two-dimensional electron gas is obtained by thermal diffusion into the semiconductor of a metallic multilayer of nickel (30~nm), gold (120~nm) and germanium (60~nm), see e.g. Ref.~\citenum{Goktas2008}.
See Methods in Ref.~\citenum{Iftikhar2015} for the estimation of the typical energy spacing between electronic levels in the central metallic island on the same sample.\\

{\noindent\textbf{Experimental setup.}} 
The measurements were performed in a dilution refrigerator including multiple filters along the electrical lines and two shields at the mixing chamber.
Conductance measurements were carried out by standard lock-in techniques at low frequencies, below 100~Hz, taking advantage of the chiral current propagation in the quantum Hall regime (see Extended Data Figure~1).
Noise measurements for the electronic temperature were performed in the MHz range using a homemade cryogenic amplifier (for details, see the supplementary information of Ref.~\citenum{Jezouin2013b}).\\

{\noindent\textbf{Electronic temperature.}} 
The displayed electronic temperatures correspond to those extracted on-chip using either quantum shot noise primary thermometry \cite{Spietz2003} or thermal noise thermometry, with error bars encapsulating also the outcome of Coulomb blockade oscillations primary thermometry (at $T\leq32$~mK) and/or standard thermometry from RuO$_2$ resistors thermally anchored to the mixing chamber (at $T\geq32$~mK).\\

{\noindent\textbf{Interface metallic `island' - 2DEG.}} 
A 2DEG-metallic island transmission probability $\tau_{\Omega-\mathrm{out}}>0.9995$ is obtained with the self calibrated procedure described below.
Here, the switches are set in open positions as in Fig.~1b (with edge channels following the red lines shown Fig.~1a and Extended Data Figure~1).
First, QPC$_{L,R}$ are set at $\tau_{L,R}=1$, in the middle of the very flat and broad intermediate plateau (thanks to the robust quantum Hall effect), and we measure the reflected signal $V_{RR}^{\tau_{L,R}=1}$ (see Extended Data Figure~1). 
The average transmission probability $\tau_{\Omega-\mathrm{out}}$ of the first (outer edge quantum Hall) channel emitted from QPC$_{L}$ and QPC$_{R}$ into the metallic island then reads:
\begin{equation}
V_{RR}^{\tau_{L,R}=1} =G_R(1-\tau_{\Omega-\mathrm{out}}/4)V_R,\nonumber
\end{equation}
with $V_R$ the (a.c.) voltage applied at the input of QPC$_R$ (see Extended Data Figure~1) and $G_R$ the gain of amplification chain $R$.
Second, we eliminate calibration uncertainties by measuring the reflected signal $V_{RR}^{\tau_{L,R}=0}=G_RV_R$ with QPC$_{L,R}$ depleted ($\tau_{L,R}=0$).
The ratio $V_{RR}^{\tau_{L,R}=1}/V_{RR}^{\tau_{L,R}=0}$ gives $\tau_{\Omega-\mathrm{out}}$ directly.
With this approach, we obtain $|1-\tau_{\Omega-\mathrm{out}}|< 5\times10^{-4}$ ($\tau_{\Omega-\mathrm{out}}\simeq 0.9997\pm0.0002$).
The same approach including also the second (inner edge quantum Hall) channel gives $\tau_{\Omega-\mathrm{in}}\simeq 0.9976$.
Note that it is usual to have better ohmic contacts with the outer quantum Hall channels, that are closest to the sample edges.\\

{\noindent\textbf{Short-circuit switch operation.}}
In practice, closing the short-circuit switches is realized by changing the voltage applied to the adjacent characterization gate (blue in Fig.~1a, see Extended Data Figure~2a for the conductance versus gate voltage of switch $R$) from $-0.35$~V (2DEG depleted/switch open) to $0.1$~V (two edge channels perfectly transmitted/switch closed).\\

{\noindent\textbf{Capacitive crosstalk corrections.}}
The transmission probability across each QPC is slightly modified when changing the voltage applied either to its adjacent characterization gate or to the gate tuning the other QPC.
Thanks to the large, micron-scale, distances this modification remains relatively small, particularly near the ballistic critical point ($<1\%$ for $\tau_{L,R}\in[0.9,1]$ when changing the adjacent switch from closed to open).
Let us first consider the crosstalk from one QPC to the other, which is more straightforward to extract. 
For this purpose, the characterization gate adjacent to the QPC for which the crosstalk is to be compensated is set to its short-circuit/closed position (as in Fig.~1c), such that changing the gate voltage tuning the other QPC is felt only through the capacitive crosstalk.
We find that this crosstalk can be precisely compensated by a relatively small shift ($\simeq-1\%$) of the split gate voltage.
Regarding now the capacitive crosstalk due to the adjacent characterization gate, the difficulty is to isolate this contribution from changes in the Coulomb blockade renormalization of the QPC conductance.
In order to suppress this renormalization, the other QPC is set in the middle of its $\tau_{L(R)}=1$ plateau and we apply a large dc bias voltage compared to the charging energy.
Extended Data Figure~2b displays the differential conductance of QPC$_R$, measured in the presence of the applied bias $V_{R}=72~\mu\mathrm{V_{dc}}$, versus gate voltage $V^\mathrm{qpc}_R$ for the adjacent switch set to position open (red line) and closed (blue line).
The gate voltage shift $\Delta V^\mathrm{ct}_R$ needed to compensate the crosstalk is determined at low QPC conductances $G^\mathrm{qpc}_R\lesssim0.1~e^2/h$, for which the dc voltage drop across the QPC is nearly independent of the switch position.
Extended Data Figure~2c displays as symbols the crosstalk compensation for QPC$_R$ in response to increasing the adjacent characterization gate voltage from $V^\mathrm{sw}_R=-0.5~$V.
The amplitude of the negative crosstalk compensation is found to increase linearly, with different slopes for different values of the switch conductance $G^\mathrm{sw}_R$.
Indeed, the capacitive crosstalk depends on the precise paths of the edge channels, which screen the gates potentials.
The crosstalk compensations used in the experiment when setting the adjacent switch from open to closed are $\Delta V_R^\mathrm{qpc}\simeq-6$~mV for QPC$_R$ and $\Delta V_L^\mathrm{qpc}\simeq-10$~mV for QPC$_L$.\\

{\noindent\textbf{Calibrations.}} 
The reflected signal $V_{RR}$ is normalized by the signal $V_{RR}^{\tau_{L,R}=0}$ measured when setting $\tau_{L,R}=0$. 
The injection voltage and amplifier gain thereby cancel out in the expression of the SET conductance $G_\mathrm{SET}$: 
\begin{equation}
G_\mathrm{SET}=\frac{2e^2}{h}(1-V_{RR}/V_{RR}^{\tau_{L,R}=0}).
\label{eqGsetr}
\end{equation}
In order to reduce the noise level, we also extract $G_\mathrm{SET}$ from the (redundant) transmitted signal $V_{LR}$ (see Extended Data Figure~1):
\begin{equation}
G_\mathrm{SET}=\frac{2e^2}{h}(V_{LR}/V_{RR}^{\tau_{L,R}=0})G_R/G_L,
\label{eqGsett}
\end{equation}
with $G_R$ ($G_L$) the gain of amplification chain $R$ ($L$).
The ratio $G_R/G_L$ is determined by setting QPC$_{L,R}$ at $\tau_{L,R}=1$ and measuring both the signals reflected ($V_{RR}^{\tau_{L,R}=1}$) and transmitted ($V_{LR}^{\tau_{L,R}=1}$):
\begin{equation}
G_R/G_L=(1-V_{RR}^{\tau_{L,R}=1}/V_{RR}^{\tau_{L,R}=0})/(V_{LR}^{\tau_{L,R}=1}/V_{RR}^{\tau_{L,R}=0})\simeq1.0105.\nonumber
\label{eqGRsGL}
\end{equation}\\

{\noindent\textbf{Experimental determination of $\Delta Q$.}}
(\textbf{i}) For $\tau_R\leq0.99$, the signal-to-noise ratio is always sufficient to accurately extract the values of $G_\mathrm{SET}^\mathrm{max,min}$ directly from the periodic conductance maximums and minimums, which stand out very strongly from the background noise.
The error bars on the visibility $\Delta Q \equiv (G_\mathrm{SET}^\mathrm{max}-G_\mathrm{SET}^\mathrm{min})/(G_\mathrm{SET}^\mathrm{max}+G_\mathrm{SET}^\mathrm{min})$ were calculated from the statistical uncertainty on $G_\mathrm{SET}^\mathrm{max,min}$, which is estimated from typically 10 different sweeps of one period.
Note that in this regime ($\tau_R\leq0.99$), the calculated error bars are found smaller than symbols size and therefore not shown.
(\textbf{ii}) For $\tau_R\in]0.99,0.998]$, although the periodic oscillations can still be clearly distinguished on the raw data (see Extended Data Figure~3), the above direct procedure would result in uncertainties that can become quite large, especially at base temperature and in the presence of a weakly transmitted second channel ($\tau_L=0.075$). 
In order to improve our extraction of $\Delta Q$, we take advantage of the observation that the conductance oscillations are sinusoidal for $\tau_R\geq0.98$ (see Extended Data Figure~3), as also expected from theory (see Eqs.~\ref{eqGsetasymqu} and \ref{eqGsetFM}, and continuous lines in Extended Data Figure~3):
The visibility of the conductance oscillations $\Delta Q$ is then extracted from a sinusoidal fit of the conductance sweeps $G_\mathrm{SET}(V_\mathrm{g})$.
The displayed error bars are the statistical error on the mean value obtained from the distinct $\Delta Q$ values obtained by fitting separately $\sim6$ different conductance sweeps.
Note that the two procedures give the same value of $\Delta Q$ in the intermediate regime $\tau_R\in[0.98,0.99]$ where they both accurately apply.
(\textbf{iii}) For $\tau_R\geq1$, there are no periodic oscillations directly visible in the raw conductance sweeps $G_\mathrm{SET}(V_\mathrm{g})$ (see right panel in Fig.~2a).
In order to put experimental bounds on the basic statement $\Delta Q\simeq0$, we have determined the visibility $\Delta Q$ (displayed Fig.~2b) using the following procedure:
First, we determine the most probable positions of the conductance maximums and minimums by `fitting' a conductance sweep (extending over typically 10 Coulomb oscillations periods) with a sinusoidal function at the known period of Coulomb oscillations, using its phase as a fitting parameter.
For each of these positions, a different value of $G_\mathrm{SET}^\mathrm{max}$ or $G_\mathrm{SET}^\mathrm{min}$ is obtained by averaging the data over an extension of one quarter of a period (assuming sinusoidal oscillations, this would result in a visibility reduction smaller than 10\%). 
Extracting separately $G_\mathrm{SET}^\mathrm{max,min}$ for the $\sim10$ periods, we can calculate their mean values and estimate the corresponding standard errors.
The error bars displayed Fig.~2b are the standard error on the mean value of $\Delta Q$, obtained from the statistical uncertainty on $G_\mathrm{SET}^\mathrm{max,min}$.\\

{\noindent\textbf{Predictions in the quantum asymmetric regime ($k_BT\ll E_C$, $\tau_L\ll1$, $1-\tau_{R}\ll1$)}}.
The conductance reads (Eq.~34 in Ref.~\citenum{Matveev2002}):
\begin{multline}
G_\mathrm{SET}^\mathrm{\tau_L\ll1,1-\tau_{R}\ll1}=\tau_L\frac{e^2}{h}\frac{2\pi^4(k_BT)^2}{3\gamma^2 E_C^2}\\
\times\left[1-2\gamma\xi\sqrt{1-\tau_R}\cos(2\pi\delta V_g/\Delta)\right],
\label{eqGsetasymqu}
\end{multline}
with $\gamma\simeq\exp(0.5772)$, $\xi\simeq1.59$, $\Delta$ the gate voltage period and $\delta V_\mathrm{g}$ the gate voltage difference from charge degeneracy.
Note that in the ballistic limit ($1-\tau_R=0$) the conductance does not depend on gate voltage but vanishes as $T^2$ following quantitatively, with the exact same prefactor, the dynamical Coulomb blockade predictions\cite{Ingold1992} for the same $E_C$ and the corresponding series resistance $R=h/e^2$.
Using Eq.~\ref{eqGsetasymqu}, the visibility of the oscillations of conductance reads:
\begin{equation}
\Delta Q(\mathrm{\tau_L\ll1,1-\tau_{R}\ll1})=2\gamma\xi\sqrt{1-\tau_R}.
\label{eqVisasymqu}
\end{equation}
Note that the temperature dependence of $G_\mathrm{SET}^\mathrm{\tau_L\ll1,1-\tau_{R}\ll1}$ (associated with dynamical Coulomb blockade) cancels out in $\Delta Q$.
Charge discreteness also affects the gate voltage dependence of thermodynamic quantities, such as the average charge ($\langle Q\rangle$) or the differential capacitance ($C_\mathrm{diff}\equiv \partial\langle Q\rangle/\partial V_\mathrm{g}$).
The effect of Coulomb blockade on thermodynamic quantities was studied most comprehensively for tunnel junctions \cite{Korshunov1987,Schon1990}: at $T=0$ and $G\gg e^2/h$, the amplitude of average charge oscillations decays exponentially with $Gh/e^2$, see e.g. Refs.~\citenum{Panyukov1991,Wang1996,Lukyanov2007}.
The theoretical extension to multi-channel junctions of arbitrary transmission, beyond the tunnel limit, was performed in Ref.~\citenum{Nazarov1999}. 
In the presence of a single nearly ballistic channel, the bosonisation approach allows for an exact solution of the average charge in the metallic island in the low energy `quantum' regime $k_BT\ll E_C$ (Eq.~26 in Ref.~\citenum{Matveev1995}):
\begin{multline}
\langle Q\rangle^\mathrm{\tau_L\ll1,1-\tau_{R}\ll1}=e V_\mathrm{g}/\Delta - (e\gamma/\pi)\sqrt{1-\tau_R}\sin (2\pi V_\mathrm{g}/\Delta)\\
+Q_0,
\label{eqQasymqu}
\end{multline}
with $Q_0$ a charge offset.
In the ballistic limit ($1-\tau_R=0$) the charge increases linearly with gate voltage, corresponding to an absence of charge quantization.
The degree of charge quantization can be characterized by the relative amplitude of the oscillations of charge or, equivalently, by the visibility of the differential capacitance ($C_\mathrm{diff}\equiv \partial\langle Q\rangle/\partial V_\mathrm{g}$) oscillations:
\begin{equation}
\Delta C_\mathrm{diff}(\mathrm{\tau_L\ll1,1-\tau_{R}\ll1})\equiv \frac{C^\mathrm{max}_\mathrm{diff}-C^\mathrm{min}_\mathrm{diff}}{C^\mathrm{max}_\mathrm{diff}+C^\mathrm{min}_\mathrm{diff}}=2\gamma\sqrt{1-\tau_R}.
\label{eqCvisasymqu}
\end{equation}
The degree of charge quantization vanishes as $\sqrt{1-\tau_R}$ when approaching the ballistic limit, and does not depend on temperature in the quantum regime ($k_BT\ll E_C$).
Importantly, the visibility in the SET conductance oscillations is directly proportional to the visibility of the differential capacitance oscillations\cite{Yi1996}, up to the fixed numerical factor $\xi\simeq 1.59$:
\begin{equation}
\Delta Q(\mathrm{\tau_L\ll1,1-\tau_{R}\ll1})=\xi\Delta C_\mathrm{diff}(\mathrm{\tau_L\ll1,1-\tau_{R}\ll1}).
\label{eqRelVisasymqu}
\end{equation}\\

{\noindent\textbf{Predictions in the quantum near ballistic regime ($k_BT\ll E_C$, $1-\tau_{L,R}\ll 1$).}} 
The conductance $G_\mathrm{SET}$ reads (Eqs.~38 and 26 in Ref.~\citenum{Furusaki1995b}):
\begin{equation}
G_\mathrm{SET}  =  \frac{e^2}{2 h} \Big[ 1 -\int^{\infty}_0 \frac{\Gamma^2_-/\cosh^2(x)}{(x\pi^2k_BT/\gamma E_C)^2+\Gamma^2_-}\mathrm{d}x \Big], \label{eqGsetFM}
\end{equation}
with $\gamma \simeq \exp(0.5772)$ and
\begin{equation}
\Gamma_{-} = (1-\tau_L)+(1-\tau_R) - 2 \sqrt{(1-\tau_L)(1-\tau_R)} \cos(2 \pi \delta V_\mathrm{g}/\Delta),\nonumber
\end{equation}
with $\Delta$ the gate voltage period.
The quantitative $\Delta Q$ predictions calculated with Eq.~\ref{eqGsetFM} are displayed as colored continuous lines in Fig.~3.
When approaching the ballistic critical point ($\sqrt{1-\tau_{L,R}}\ll k_BT/E_C\ll1$), the visibility $\Delta Q$ reduces to the simple asymptotic expression (Eq.~2 in main text):
\begin{equation}
\Delta Q(\sqrt{1-\tau_{L,R}}\ll\frac{k_BT}{E_C}\ll1)=\frac{\gamma E_C}{\pi k_BT}\sqrt{(1-\tau_L)(1-\tau_R)}. 
\label{eqVisFMlimbal}
\end{equation}
The differential capacitance ($C_\mathrm{diff}$) when one QPC approaches the ballistic critical point ($\tau_R\rightarrow1$) reduces to the asymptotic expression (Eq.~41 in Ref.~\citenum{LeHur2002}):
\begin{multline}
C_\mathrm{diff}^{1-\tau_R\ll1-\tau_L\ll1}=-4\gamma(e/\Delta)\ln (1-\tau_L)\sqrt{(1-\tau_L)(1-\tau_R)}\\
\times\cos\left(\frac{2\pi\delta V_\mathrm{g}}{\Delta}\right)+\frac{e}{\Delta}\, ,
\label{eqCdiffLSlimbal}
\end{multline}
and the visibility in the oscillations of the differential capacitance reads:
\begin{equation}
\Delta C_\mathrm{diff}(1-\tau_R\ll1-\tau_L\ll1)=-4\gamma\ln (1-\tau_L)\sqrt{(1-\tau_L)(1-\tau_R)}.
\label{eqVisCdiffLSlimbal}
\end{equation}
We recover the same $\sqrt{1-\tau_R}$ scaling behavior near the ballistic critical point ($\tau_R=1$) that was found in the asymmetrical regime (Eqs.~\ref{eqCvisasymqu} and \ref{eqVisasymqu}), and which is also found in the visibility of the conductance Coulomb oscillations (Eq.~\ref{eqVisFMlimbal}).
Note that for two identical (e.g. spin-degenerate) channels ($\tau\equiv\tau_L=\tau_R$) near the ballistic critical point ($1-\tau\ll1$), the differential capacitance reads (Eqs.~49 and 52 in Ref.~\citenum{Matveev1995}, a factor $e\Delta/2E_C$ was applied to match the definition $C_\mathrm{diff}\equiv \partial\langle Q\rangle/\partial V_\mathrm{g}$):
\begin{multline}
C_\mathrm{diff}^{1-\tau\equiv1-\tau_L=1-\tau_R\ll1}=\frac{4\gamma e}{\pi\Delta}\ln\left[(1-\tau)\sin^2\left(\frac{\pi\delta V_\mathrm{g}}{\Delta}\right)+\frac{k_BT}{E_C}\right]\\
\times(1-\tau)\cos\left(2\pi\delta V_\mathrm{g}/\Delta\right)+e/\Delta.
\label{eqCdiffdeg}
\end{multline}
When approaching the ballistic critical point ($\tau\rightarrow1$), the visibility in the oscillations of the differential capacitance therefore asymptotically vanishes as $1-\tau$, as in Eq.~\ref{eqVisFMlimbal} with $\tau_L=\tau_R$.
\\

{\noindent\textbf{Predictions in the presence of strong thermal fluctuations ($k_BT\gg E_C/\pi^2$).}} 
Charge discreteness leads to periodic oscillations of the observables ({\it e.g.}, conductance and differential capacitance) while sweeping a capacitively coupled gate voltage.
Quantum fluctuations decrease the oscillations, which are further attenuated by thermal fluctuations for increasing temperature, until the amplitude becomes exponentially small for $k_BT\gg E_C/\pi^2$.
The exponential temperature dependence in $k_BT/E_C$ is quite robust, applying both to thermodynamic \cite{Korshunov1987,Schon1990,Nazarov1999} and transport (Methods) properties.
It can be demonstrated in the limits both of small and large transmission probabilities of the conduction channels comprising the junctions, and for various models of the metallic island.
Remarkably, the presence of thermal fluctuations not only preserve the quantum $\sqrt{1-\tau}$ suppression of the oscillations, but it is expected from the results of Ref.~\citenum{Nazarov1999} that the square root scaling of the differential capacitance extends with increasing temperature, up to the full range of $\tau_{L,R}\in[0,1]$.
Once again, we note that the relative oscillations in the differential capacitance and in the conductance characterize equally well the degree of charge quantization, both following the same $\sim\exp \left[ -\pi^2k_BT/E_C\right]\sqrt{(1-\tau_L)(1-\tau_R)}$ behavior.
Further information regarding the predictions and theoretical methods in the presence of strong thermal fluctuations are provided in the four following sections.
\\

{\noindent\textbf{Differential capacitance in the tunnel limit with strong thermal fluctuations ($k_BT\gg E_C/\pi^2$, $\tau_{L,R}\ll 1$).}}
To start with, we evaluate the oscillatory part of the island's free energy in the limit $\tau_{L,R}\ll 1$, where the suppression of charge quantization is entirely due to thermal fluctuations. 
Considering high temperatures, it is convenient to transform the isolated island's partition function,
\begin{equation}
Z=\!\sum_{n=-\infty}^{\infty}\!\exp\left\{-\frac{E_n({\cal N})}{k_BT}\right\}\,,\,\, E_n({\cal N})=E_C(n-{\cal N})^2\,,
\nonumber
\end{equation}
using the Poisson summation formula; the result is
\begin{equation}
Z=\sqrt{\frac{\pi k_BT}{E_C}}\!\sum_{k=-\infty}^{\infty}\!e^{-2\pi i k{\cal N}} \exp\left\{-\frac{\pi^2k_BT}{E_C}k^2\right\}\,.
\label{ZPoiss}
\end{equation}
Here ${\cal N}\equiv V_g/\Delta$ (with $\Delta$ the period in gate voltage $V_g$) is the charge induced by the gate voltage in units of $e$, and the summations are performed over integer $n$ and $k$. 
The $k=0$ and $k=\pm 1$ terms in the sum of Eq.~\ref{ZPoiss} yield, respectively, the leading ${\cal N}$-independent and ${\cal N}$-dependent contributions $F_0$ and $\delta F({\cal N})$ to the free energy $F=-k_BT\ln Z$ at $k_BT\gg E_C/\pi^2$.
The resulting oscillatory part of the differential capacitance,
\begin{align}
C_\mathrm{diff}^{\tau_{L,R}\ll1}\equiv&\frac{e}{\Delta}\left(1-\frac{1}{2E_C}\partial^2_{\cal N}F\right)\nonumber\\
=&\frac{e}{\Delta}-4\frac{e}{\Delta}\frac{\pi^2k_BT}{E_C} \exp\left\{-\frac{\pi^2k_BT}{E_C}\right\}\cos(2\pi{\cal N})\,,
\label{diffC1}
\end{align}
is exponentially suppressed at high temperatures. \\

{\noindent\textbf{Differential capacitance in the near ballistic regime with strong thermal fluctuations ($k_BT\gg E_C/\pi^2$, $1-\tau_{R}\ll1$).}}
A similar suppression of oscillations of the thermodynamic characteristics can also be demonstrated in the case of high-transmission junctions, where both thermal and quantum fluctuations contribute to the reduction of charge quantization.
For definiteness, we consider here a single-junction case ($\tau_L=0$) with $1-\tau_R\ll1$.
Evaluation of $C_\mathrm{diff}^{\tau_L=0,1-\tau_R\ll1}$ can be performed using the bosonization scheme developed in Ref.~\citenum{Matveev1995}.
In that formalism, the $\cal N$-dependent part of the differential capacitance reads $\delta C_\mathrm{diff}^{\tau_L=0,1-\tau_R\ll1}=-(e/\Delta)(2\pi^2/E_C)D\sqrt{1-\tau_R}\langle\cos[2\pi{\cal N}-\varphi(0)]\rangle$, where the bosonic quantum field $\varphi (0)=2\pi{\hat Q}/e$ corresponds to the charge $\hat Q$ passed through the junction ($x=0$), and $D$ is the energy bandwidth appearing in the definition of boson variables.
Averaging $\langle\dots\rangle$ is performed over the fluctuations of the field $\varphi (x)$.
The Hamiltonian describing these fluctuations consists of two parts~\cite{Matveev1995}, representing, respectively, the energy of particle-hole excitations and the charging energy.
The former part depends on $(\nabla\varphi)^2$, while the latter one has the form $E_C[\varphi(0)/2\pi]^2$.
Replacement of the ground-state averaging~\cite{Matveev1995} with an average over the Gibbs distribution of fluctuations, $\propto\exp\{-(E_C/k_BT)[\varphi(0)/2\pi]^2\}$, results in the renormalization of the bandwidth $D$ to a physically meaningful value $\sim k_BT$, and in exponential suppression of the oscillations at $k_BT\gg E_C/\pi^2$:
\begin{multline}
\delta C_\mathrm{diff}^{\tau_L=0,1-\tau_R\ll1}
\sim-4\frac{e}{\Delta}\frac{\pi^2k_BT}{E_C}\exp\left\{-\frac{\pi^2k_BT}{E_C}\right\}\sqrt{1-\tau_R}\\
\times\cos(2\pi{\cal N})
\label{diffC2}\,.
\end{multline}
As it follows from Ref.~\citenum{Nazarov1999}, Eq.~\ref{diffC2} is applicable in the full range of $\tau_R$ for $k_BT\gg E_C/\pi^2$ (the numerical coefficient in Eq.~\ref{diffC2} was established with the help of Ref.~\citenum{Nazarov1999}).
The identical exponential suppression for an almost-isolated island (Eq.~\ref{diffC1}) is therefore simply the limit case $\tau_R\ll1$ of Eq.~\ref{diffC2}. 
In addition, quantum fluctuations contribute to the same suppression factor $\sqrt{1-\tau_R}$ derived at $1-\tau_R\ll1$ in the quantum regime $k_BT\ll E_C$ (Eq.~\ref{eqCvisasymqu}).
Furthermore, Eq.~\ref{diffC2} derived for $k_BT\gg E_C$ matches the $T=0$ result of Ref.~\citenum{Matveev1995} at $k_BT\sim E_C$; given the large numerical factor $\pi^2$ in the exponent of Eq.~\ref{diffC2}, there may be, however, a broad crossover temperature region between the two limits.\\

{\noindent\textbf{Conductance in the tunnel limit with strong thermal fluctuations ($k_BT\gg E_C/\pi^2$, $\tau_{L,R}\ll1$).}}
Turning now to conductance oscillations, we again start from the simpler case of  low-transmission barriers ($\tau_{L,R}\ll1$).
In that limit, the rate equation for current carried by spin-polarized electrons yields~\cite{Glazman1989}:
\begin{multline}
G_{\rm SET}^{\tau_{L,R}\ll1}({\cal N},T)=\frac{e^2}{h}\frac{\tau_L\tau_R}{\tau_L+\tau_R}
\sum_{n=-\infty}^\infty\frac{\exp\left[-E_n({\cal N})/k_BT\right]}{Z({\cal N},T)}\\
\times f\left(\frac{E_n({\cal N})-E_{n-1}({\cal N})}{k_BT}\right)\,,
\label{G1}
\end{multline}
where $f(x)=x/(1-e^{-x})$. Application of the Poisson summation formula to Eq.~\ref{G1} is tedious but straightforward. 
The result is an expression for $G_{\rm SET}^{\tau_{L,R}\ll1}$ involving a sum of harmonics $\sim\cos(2\pi k{\cal N})$, similar to Eq.~\ref{ZPoiss}. 
The largest term,
\begin{equation}
G_\infty=\frac{e^2}{h}\left[\frac{1}{\tau_L}+\frac{1}{\tau_R}\right]^{-1}\,,
\label{G0}
\end{equation}
does not oscillate and is simply the conductance of two resistors connected in series. The leading oscillatory term, 
\begin{equation}
\delta G_{\rm SET}^{\tau_{L,R}\ll1}({\cal N},T)=-G_\infty\frac{\pi^2k_BT}{E_C}
\exp\left\{-\frac{\pi^2k_BT}{E_C}\right\}\cos(2\pi{\cal N})\,,
\label{deltaG1}
\end{equation}
exhibits the same exponential suppression as the differential capacitance (Eq.~\ref{diffC1}).\\

{\noindent\textbf{Conductance in the near ballistic regime with strong thermal fluctuations ($k_BT\gg E_C/\pi^2$, $1-\tau\ll1$).}}
Regarding now the conductance across a metallic island with high-transmission contacts, we first (\textbf{A}) present a formalism somewhat different from Ref.~\citenum{Furusaki1995b}, details of which will be published separately \cite{Idrisov2016}, and then (\textbf{B}) further establish the predictions by extending the formalism of Ref.~\citenum{Furusaki1995b} to high temperatures.

{\noindent (\textbf{A})} In the first approach, we start from the chiral edge excitations of the integer quantum Hall regime, in close correspondence with the experimental configuration.
Note that although we are interested in the high-temperature limit, all the energy scales in the experiment remain much smaller than the quantum Hall energy gap.  
At such low energies, the quantum Hall edge states may be described by the effective theory \cite{Halperin1982,Wen1990,Frohlich1991}.
According to this theory, edge excitations can be viewed as bosonic edge magneto-plasmons.
The corresponding one-dimensional charge density waves $\rho_{s\alpha}(x)$ ($s\in\{L,R\}$, $\alpha\in\{1,2\}$, see Extended Data Figure~4 for notations) verify the canonical commutation relations
$[\rho_{s\alpha}(x), \rho_{s'\beta}(y)] = (-1)^\alpha2\pi i e^2\delta_{ss'}\delta_{\alpha\beta}\delta'(x-y)$, where the sign accounts for the propagation direction of the chiral edge states.
The Hamiltonian of the experimental setup contains three terms, ${\cal H} = {\cal H}_0 + {\cal H}_{\rm int} + {\cal H}_T$. 
The first term describes the bare edge states dynamics ${\cal H}_0 = (hv_F/2e^2)\sum_{s\alpha}\int dx \rho^2_{s\alpha}(x)$.
The second term describes Coulomb interactions at the metallic island:
\begin{align}
\label{hint}
{\cal H}_{\rm int} &=E_C({\hat Q}/e-{\cal N})^2,\\
\label{charge}
{\hat Q}&=\frac{e}{2\pi}[\varphi_L(0)+\varphi_R(0)]\nonumber\\
&=\sum_\alpha\big[\int_0^\infty\!\! dx \rho_{L\alpha}(x) + \int_{-\infty}^0\!\! dx \rho_{R\alpha}(x)\big]\,.
\end{align}
Note that the first equality in Eq.~\ref{charge} defines the Bose field operators also used in the derivation of Eq.~\ref{diffC2}, but here for the case of two contacts.
The last term describes the backscattering of electrons at the two QPCs:
\begin{align}
{\cal H}_T &= (A_L + A_R + {\rm h.c.}),\\
A_s &= \gamma_s \psi^\dag_{s1}(0)\psi_{s2}(0),\label{ampl} \\
\psi_{s\alpha}(0) &= \sqrt{\frac{D}{hv_F}}\exp\left\{2\pi i\int_{-\infty}^0 \!\!\!dx \rho_{s\alpha}(x)/e\right\}\,,
\label{ampl2}
\end{align}
where the backscattering amplitudes $\gamma_{L,R}$ depend on the `intrinsic' transmission probabilities $\tau_{L,R}$ (in the near ballistic regime, $1-\tau_{L(R)}\simeq |\gamma_{L(R)}|^2/(\hbar v_F)^2$).
Note that we set to zero the distance metallic `island'-QPCs, which is much shorter then the wavelength of excitations in the experiment.
Importantly, we stress that the exact same Hamiltonian arises in the absence of the quantum Hall effect, when applying the bosonization procedure to a metallic island connected to reservoirs through spin-polarized electron channels (as in Refs.~\citenum{Matveev1995,Furusaki1995b}).
Consequently, the predictions below apply beyond the quantum Hall configuration here used as a starting point.
Now focusing on the near ballistic regime $1-\tau_{L,R}\ll1$, we apply the scattering theory approach developed in Ref.~\citenum{Slobodeniuk2013,Sukhorukov2016}.
The average $\langle I\rangle\equiv {\rm Tr}(\rho I)$ of the current operator $I = v_F[\rho_{R1}(0)-\rho_{R2}(0)]$ is evaluated perturbatively in backscattering amplitudes (Eq.~\ref{ampl}).
With this aim in view, we express the density matrix $\rho = U\rho_0 U^\dag$ in terms of its equilibrium value $\rho_0 \propto \exp[-({\cal H}_0+{\cal H}_{\rm int})/k_BT]$, and expand the evolution operator $U = {\rm \hat T exp}\big[-2\pi i\int dt{\cal H}_T(t)/h\big]$ in powers of $\gamma_s$.
This results in the two leading terms:
\begin{equation}
\label{curr}
\langle I\rangle = \langle I\rangle_0 + \frac{1}{\hbar^2}\langle \iint dt'dt''[{\cal H}_T (t'),[{\cal H}_T(t''), I]]\rangle_0,
\end{equation}
where the average is now taken with respect to the equilibrium density matrix $\rho_0$. 
Note that the Hamiltonian ${\cal H}_0+{\cal H}_{\rm int}$ is quadratic in plasmon operators.
Consequently, the corresponding dynamics can be accounted for exactly within the scattering theory approach for bosons \cite{Slobodeniuk2013,Sukhorukov2016}.
For instance, the scattering matrix for the interaction Hamiltonian ${\cal H}_{\rm int}$ (here ignoring the backscattering Hamilitonian ${\cal H}_T$), which relates the currents in the incoming $(L1, R2, L2, R1)$ and outgoing $(L2,R1,L1,R2)$ channels at the frequency $\omega/2\pi$, reads:
\begin{equation}
\label{matrix}
S(\omega) = \frac{1}{2}\left(
\begin{array}{cccc}
z & z & 2-z & -z \\
z & z & -z & 2-z \\
2-z & -z & z & z \\
-z & 2-z & z & z
\end{array}
\right),
\end{equation}
where $z = 1/(ih\omega/4E_C +1)$.
Taking the limit $\omega \to 0$ \cite{Sukhorukov2016}, one finds for the first term in Eq.~\ref{curr}, $ \langle I\rangle_0 = e^2V_\mathrm{dc}/2h$.
The bare conductance is thus half the conductance quantum.
In the limit of small dc bias $V_\mathrm{dc}$ the second term can be rewritten as
\begin{equation}
\label{tun}
\delta \langle I\rangle =(e/\hbar^2) \int dt \langle [A^\dag_L(t) + A^\dag_R(t), A_L(0) + A_R(0)]\rangle_0
\end{equation}
This term contains the coherent contribution
\begin{multline}
\delta \langle I\rangle_{\rm osc} = (e/\hbar^2) {\rm Re}\,\gamma_L^*\gamma_R \\
\times\int dt \langle \psi^\dag_{L2}(0,t)\psi_{L1}(0,t)\psi^\dag_{R1}(0,0)\psi_{R2}(0,0)\rangle_0,
\label{coh-curr}
\end{multline}
which oscillates as a function of the induced charge $e{\cal N}$.
In general, one can again use the scattering matrix Eq.~\ref{matrix} to evaluate the average in Eq.~\ref{coh-curr}, which leads to a complex expression \cite{Idrisov2016}.
However, the leading high-temperature asymptotics can be found using exactly same argument as for the case of the differential capacitance considered above.
Specifically, according to Eq.~\ref{ampl2} the particular value of the charge $Q$ in the island leads to the phase shift $e^{2\pi i (Q/e-\cal{N})}$ in the correlation function in Eq.~\ref{coh-curr}.
Therefore, by averaging the correlation function over instant fluctuations of this charge, which are distributed with the equilibrium Gibbs weights $\propto\exp [-(Q/e)^2E_C/k_BT]$, one finds the high-temperature behavior of the oscillating part of the current:
\begin{align}
\delta \langle I\rangle_{\rm osc} \propto&\frac{e^2V_\mathrm{dc}}{h}\sqrt{\frac{E_C}{k_BT}}\sqrt{(1-\tau_L)(1-\tau_R)}\nonumber\\
&\times\int dQ \exp\left\{-\frac{Q^2E_C}{e^2k_BT}\right\} \cos[2\pi({\cal N}- Q/e)]\nonumber\\
\propto&\frac{e^2V_\mathrm{dc}}{h}\exp\left\{-\frac{\pi^2k_BT}{E_C}\right\}\sqrt{(1-\tau_L)(1-\tau_R)}\cos(2\pi{\cal N}).
\end{align}
The validity of this simplified approach is confirmed by detailed calculations in Ref.~\citenum{Idrisov2016}.

{\noindent (\textbf{B})} An alternative route of calculation amounts to re-working Eq.~A5 of Ref.~\citenum{Furusaki1995b} for the case $1-\tau_{L,R}\ll1$, or Eq.~A27 for the asymmetric case $\tau_L\ll1$, $1-\tau_R\ll1$. 
In either case, the largest term in the limit $k_BT\gg E_C/\pi^2$ is, unsurprisingly, ${\cal N}$-independent.
Like Eq.~\ref{G0} above, it represents the conductance of two junctions connected in series, $G_\infty\approx e^2/2h$ in the case of $1-\tau_{L,R}\ll1$, and $G_\infty\approx (e^2/h)\tau_L$ if $\tau_L\ll 1$, $1-\tau_R\ll1$.
The leading oscillatory term in the former case is
\begin{multline}
\delta G_{\rm SET}^{1-\tau_{L,R}\ll1}({\cal N},T)\sim\frac{e^2}{h}
\exp\left\{-\frac{\pi^2k_BT}{E_C}\right\}\sqrt{(1-\tau_L)(1-\tau_R)}\\
\times\cos(2\pi{\cal N})\,.\label{deltaG2}
\end{multline}
In the asymmetric case, the factor $\sqrt{(1-\tau_L)(1-\tau_R)}$ in the above expression is replaced by $\tau_L\sqrt{1-\tau_R}$.
Regarding now the visibility of conductance oscillations, it reads:
\begin{equation}
\label{result}
\Delta Q \sim \exp\left\{-\frac{\pi^2k_BT}{E_C}\right\}\sqrt{(1-\tau_L)(1-\tau_R)}.
\end{equation}
This form correctly extrapolates between the symmetric and asymmetric cases.\\

{\noindent\textbf{Conductance at $T\simeq17$~mK vs quantum regime predictions.}} 
Although the visibility $\Delta Q$ of the oscillations in the SET conductance best reflects the degree of charge quantization, we can also confront experiment and theory directly at the underlying conductance sweeps level.
In Extended Data Figure~3, we compare $G_\mathrm{SET}(\delta V_\mathrm{g})$ measurements (symbols) and predictions near the ballistic critical point ($1-\tau_R\simeq0.02$ and $0.004$) with QPC$_L$ in both the tunnel ($\tau_L=0.075$) and almost perfectly transmitted ($1-\tau_L\simeq0.02$) regimes.
Continuous lines are calculated with the electronic temperature $T=17$~mK, using Eq.~\ref{eqGsetasymqu} for the two top panels (asymmetric regime, $\tau_L=0.075$) and Eq.~\ref{eqGsetFM} for the two bottom panels (near ballistic regime,  $\tau_L=0.983$).
The grey areas correspond to the experimental uncertainty $\pm4$~mK.
The demonstrated agreement validates the full prediction for the renormalized SET conductance.\\

{\noindent\textbf{Charge quantization criteria: conductance versus transmission probabilities.}} 
Theory predicts that as soon as one conduction channel connected to the metallic island is ballistic, the charge in the island is completely unquantized.
In the manuscript we show that charge quantization collapses systematically at the ballistic critical point $\tau_R=1$, independently of the setting of the second channel ($\tau_L<1$).
Here, we further demonstrate that the crucial ingredient is not the overall conductance but the presence of a perfectly transmitted channel.
For this purpose, we compare the two configurations displayed in Extended Data Figure~5a,b.
In both configurations, QPC$_L$ is tuned to the same standard setting corresponding to a single conduction channel of `intrinsic' transmission probability $\tau_L=0.24$.
In both configurations, QPC$_R$ is set to the same overall `intrinsic' conductance $G^\mathrm{qpc}_R\equiv\tau_Re^2/h=1.5e^2/h$.
However, in configuration (a) QPC$_R$ decomposes into one ballistic channel and one channel of `intrinsic' transmission probability $0.5$, whereas in configuration (b) it decomposes into two non-ballistic channels of `intrinsic' transmission probabilities $0.7$ and $0.8$.
(In practice, QPC$_R$ of configuration (b) is realized using two different physical QPCs biased at the same voltage.)
As shown Extended Data Figure~5c, the SET conductance displays strong oscillations in configuration (b), signaling charge quantization in the absence of a ballistic channel.
In striking contrast, the SET conductance in configuration (a) does not depend on gate voltage, signaling a completely unquantized island charge in the presence of one ballistic channel.

\begin{figure*}[p]
\renewcommand{\figurename}{\textbf{Extended Data Figure}}
\renewcommand{\thefigure}{\textbf{1}}
\centering\includegraphics [width=1\textwidth]{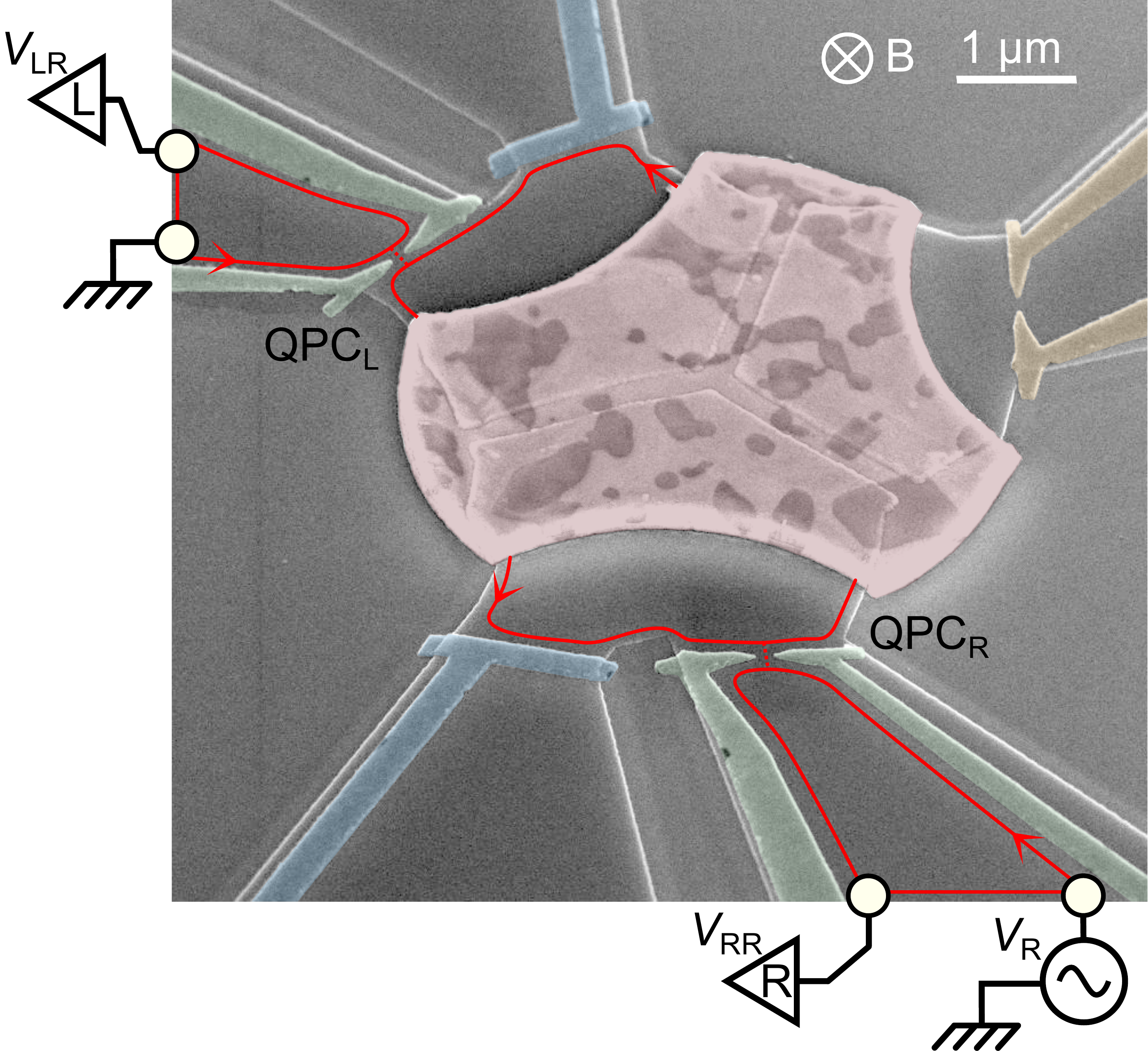}
\caption{\small
\textbf{Measurement schematic.} 
The signal $V_{LR}$ ($V_{RR}$) is the voltage measured with amplification chain $L$ ($R$) in response to the injected voltage $V_R$.
The trenches etched in the 2DEG, that can be seen in the form of a Y through the metallic island, ensure that the only way from one QPC to the other is across the metallic island.
The experiment is performed in the quantum Hall regime at filling factor $\nu=2$, where the current propagates along the edges in the direction indicated by arrows.
}
\end{figure*}

\begin{figure*}[p]
\renewcommand{\figurename}{\textbf{Extended Data Figure}}
\renewcommand{\thefigure}{\textbf{2}}
\centering\includegraphics [width=1\textwidth]{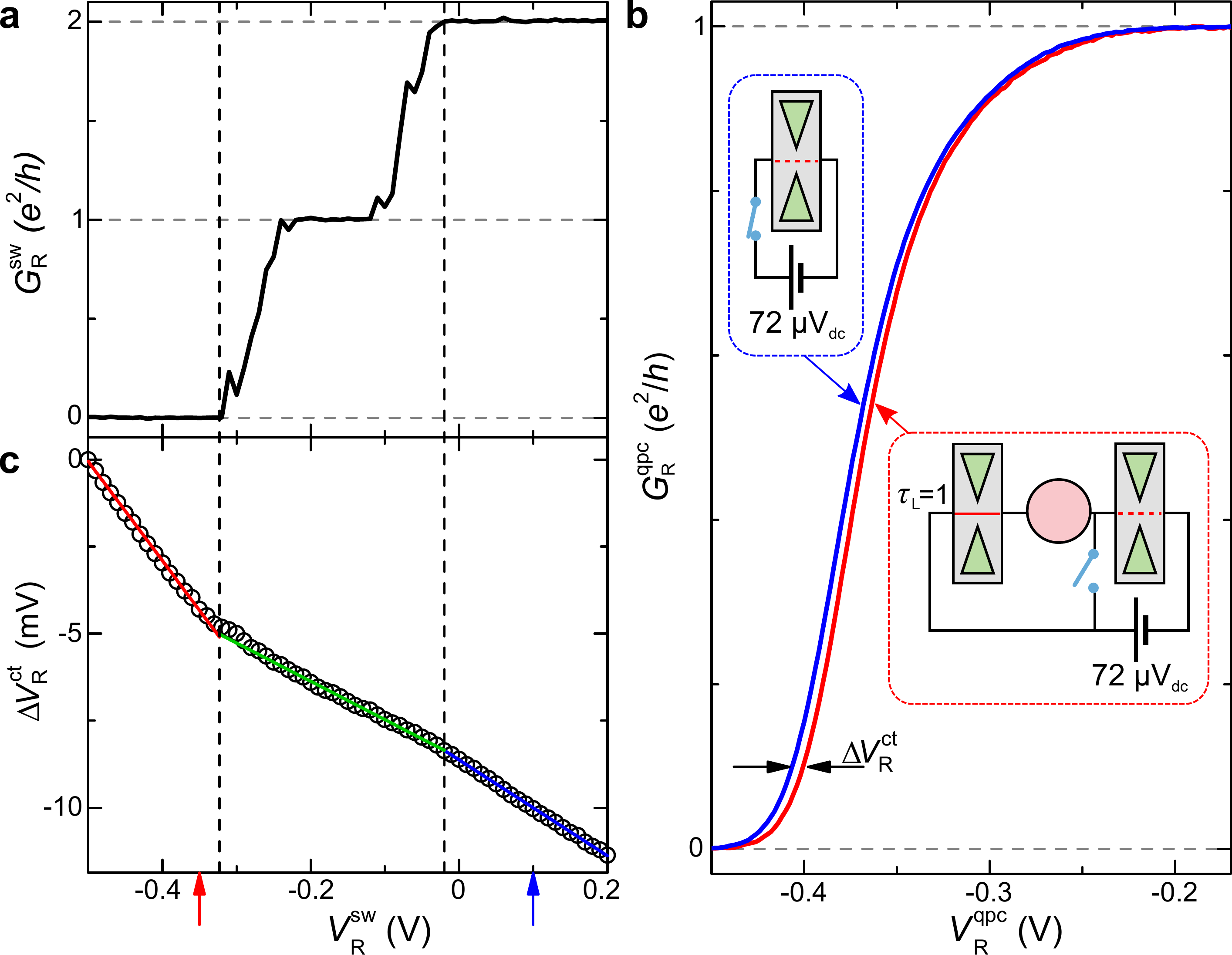}
\caption{\small
\textbf{Crosstalk compensation.} 
\textbf{a}, (Intrinsic) conductance $G^\mathrm{sw}_R$ across the characterization gate adjacent to QPC$_R$, versus gate voltage $V^\mathrm{sw}_{R}$.
In the experiment, the left and right switches are independently set to the open/closed positions with $V^\mathrm{sw}_{R,L}=-0.35$~V/$0.1$~V, respectively (vertical arrows in panel (c)).
\textbf{b}, QPC$_R$ differential conductance in the presence of $72~\mu\mathrm{V}_\mathrm{dc}$, versus QPC gate voltage $V^\mathrm{qpc}_R$.
The red/blue lines are measured with the adjacent switch in the open/closed positions, respectively.
Note that the voltage drop across QPC$_R$ is smaller with the switch open, due to the added series resistance.
Although this does not result in a large error, since $G^\mathrm{qpc}_R$ depends weakly on voltage bias, this effect is minimized by extracting the crosstalk compensation $\Delta V^\mathrm{ct}_R$ at low $G^\mathrm{qpc}_R\lesssim0.1~e^2/h$. 
\textbf{c}, Symbols represent the crosstalk compensation $\Delta V^\mathrm{ct}_R$, with respect to the gate voltage $V^\mathrm{sw}_{R}=-0.5$~V, versus $V^\mathrm{sw}_{R}$.
Lines are linear fits of the crosstalk compensation at $G^\mathrm{sw}_{R}=0$ (red, $-2.8\%$ relative compensation), $0<G^\mathrm{sw}_{R}<2e^2/h$ (green, $-1.1\%$ relative compensation) and $G^\mathrm{sw}_{R}=2e^2/h$ (blue, $-1.4\%$ relative compensation).
}
\end{figure*}

\begin{figure*}[p]
\renewcommand{\figurename}{\textbf{Extended Data Figure}}
\renewcommand{\thefigure}{\textbf{3}}
\centering\includegraphics [width=1\textwidth]{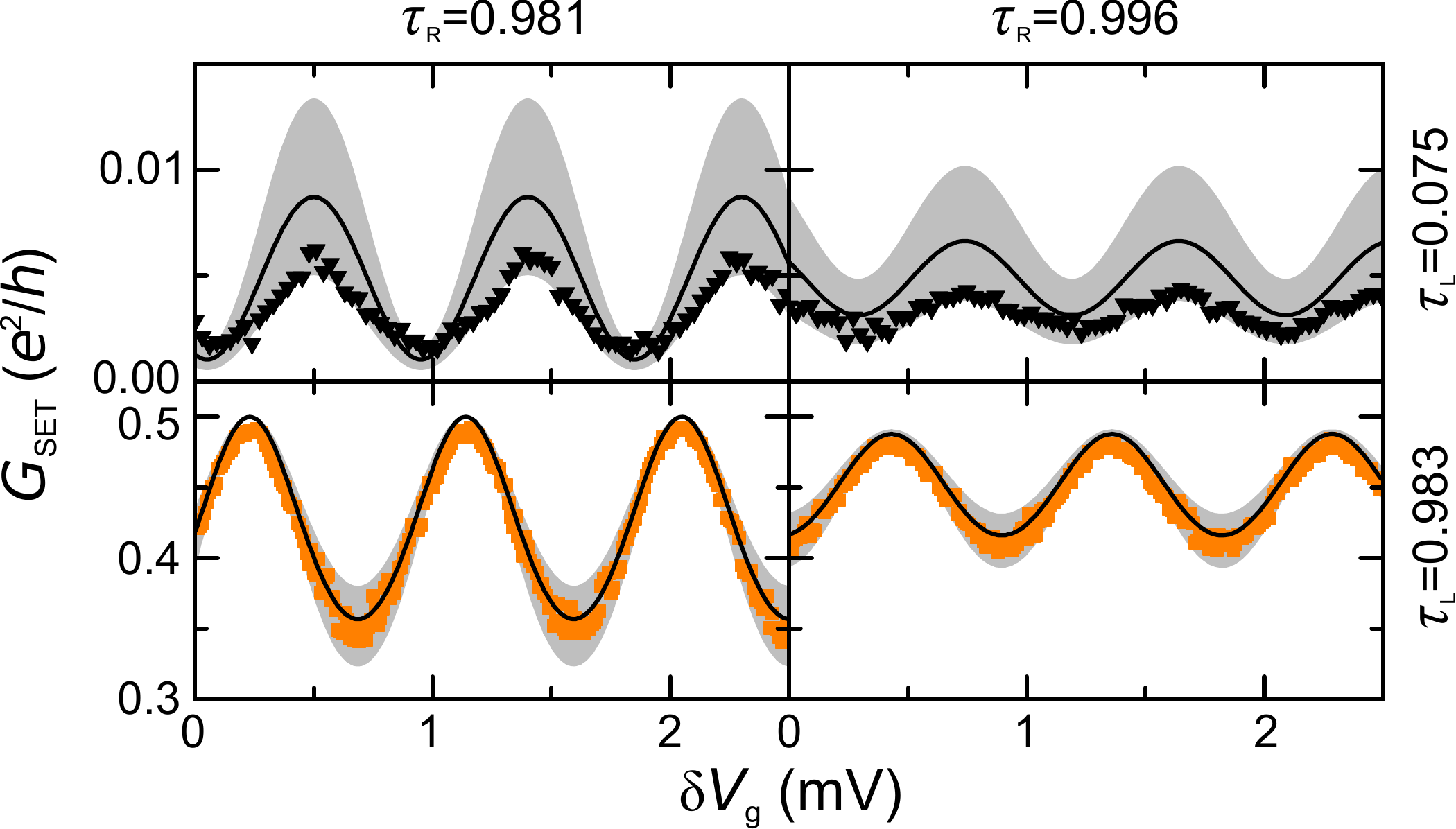}
\caption{\small
\textbf{Conductance measurements versus quantitative predictions.} 
Direct $G_\mathrm{SET}(\delta V_g)$ comparison at $T\simeq17$~mK between data (symbols) and predictions (continuous lines, grey areas correspond to the temperature uncertainty $\pm4$~mK) in the two limits addressed by theory (Eq.~\ref{eqGsetasymqu} for $\tau_L\sim0$, Eq.~\ref{eqGsetFM} for $\tau_L\sim1$). 
}
\end{figure*}

\begin{figure*}[p]
\renewcommand{\figurename}{\textbf{Extended Data Figure}}
\renewcommand{\thefigure}{\textbf{4}}
\centering\includegraphics [width=1\textwidth]{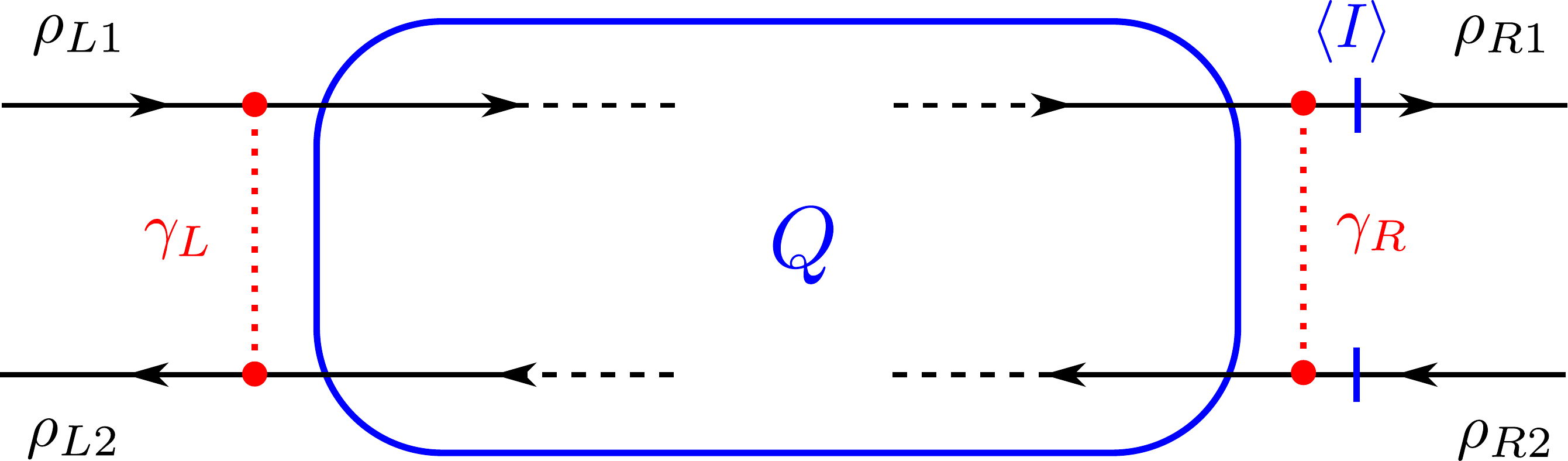}
\caption{\small
\textbf{Theoretical description of the experimental setup in formalism (A) for strong thermal fluctuations.}
We consider the regime of the quantum Hall effect, where only one spinless edge mode contributes to the transport.
The corresponding edge states are described by four charge density operators, labeled by $s\in\{L,R\}$ and $\alpha\in\{1,2\}$.
These states are mixed (backscattered) at the two QPCs (red dashed lines) with amplitudes $\gamma_L$ and $\gamma_R$ (Eqs.~\ref{ampl} and \ref{ampl2}).
The edge densities enter into the interaction Hamiltonian (Eq.~\ref{hint}) through the total charge $\hat Q$ of the metallic island (Eq.~\ref{charge}).
The average current $\langle I\rangle$ is calculated through a cross-section immediately to the right of QPC$_R$ (vertical blue lines). 
}
\label{scheme1}
\end{figure*}

\begin{figure*}[p]
\renewcommand{\figurename}{\textbf{Extended Data Figure}}
\renewcommand{\thefigure}{\textbf{5}}
\centering\includegraphics [width=1\textwidth]{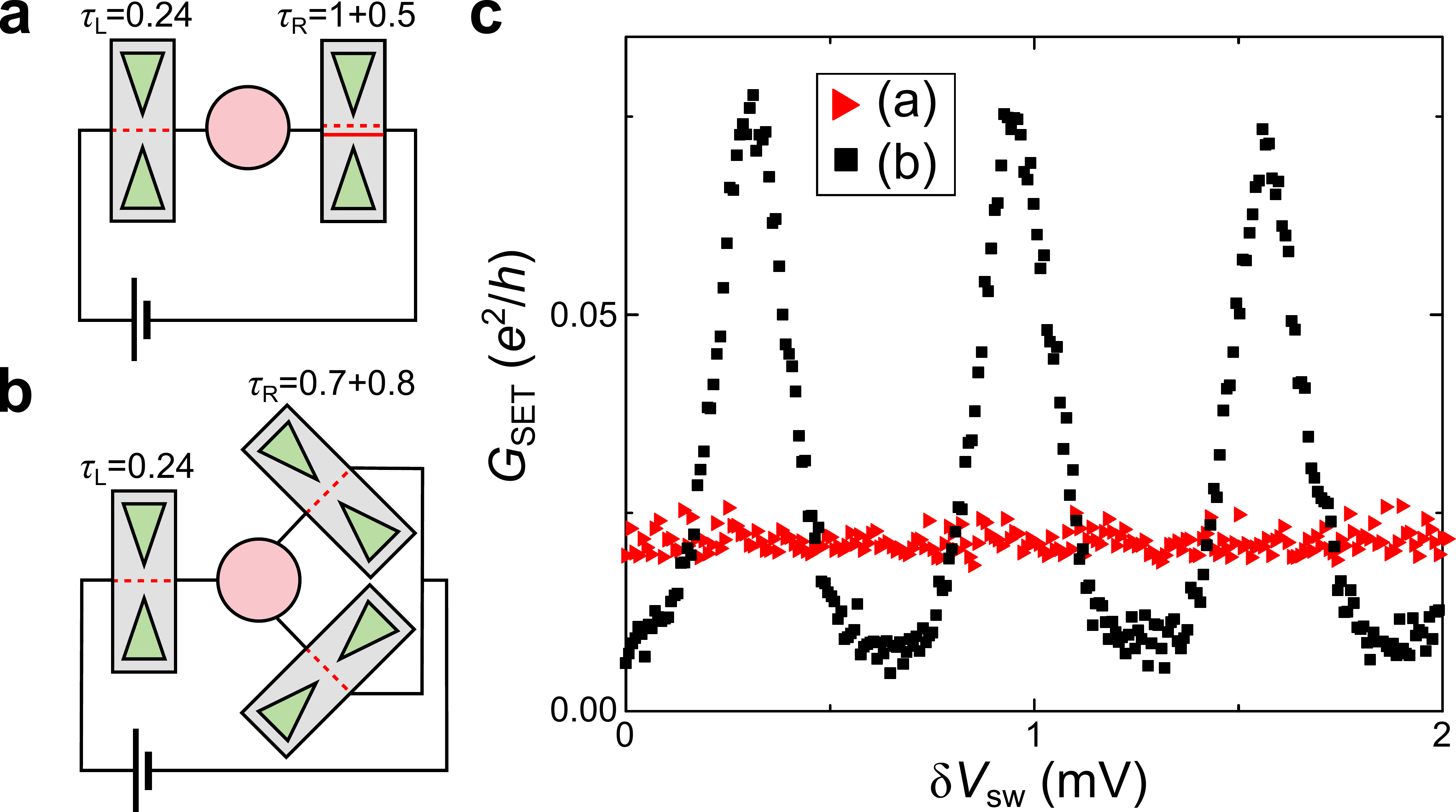}
\caption{\small
\textbf{Charge quantization: conductance vs transmission probabilities.} 
\textbf{a},\textbf{b}, Schematics of the compared configurations, both with the same QPC$_L$ setting $\tau_L=0.24$.
In configuration (a), QPC$_R$ is set to an `intrinsic' conductance $G^\mathrm{qpc}_R\equiv\tau_Re^2/h=1.5e^2/h$, that decomposes into one ballistic channel and one channel of `intrinsic' transmission probability $0.5$. 
In configuration (b), QPC$_R$ is set to the same `intrinsic' conductance $G^\mathrm{qpc}_R=1.5e^2/h$, that now decomposes into two non-ballistic channels of `intrinsic' transmission probabilities $0.7$ and $0.8$.
\textbf{c}, Sweeps of the device conductance are plotted as symbols versus gate voltage for the two configurations.
Conductance oscillations are visible only in configuration (b), in the absence of a ballistic channel connected to the island.
}
\end{figure*}


\begin{thebibliography}{44}
\expandafter\ifx\csname natexlab\endcsname\relax\def\natexlab#1{#1}\fi
\expandafter\ifx\csname url\endcsname\relax
  \def\url#1{\texttt{#1}}\fi
\expandafter\ifx\csname urlprefix\endcsname\relax\def\urlprefix{URL }\fi

\bibitem[{Sch\"on \& Zaikin(1990)}]{Schon1990}
Sch\"on, G. \& Zaikin, A.~D.
\newblock Quantum coherent effects, phase transitions, and the dissipative
  dynamics of ultra small tunnel junctions.
\newblock \emph{Phys. Rep.} \textbf{198}, 237--412 (1990).

\bibitem[{Grabert \& Devoret(1992)}]{Ingold1992}
Grabert, H. \& Devoret, M.~H. (eds.).
\newblock \emph{Single charge tunneling} (1992), plenum, new york edn.

\bibitem[{Likharev(1999)}]{Likharev1999}
Likharev, K.~K.
\newblock Single-electron devices and their applications.
\newblock \emph{Proc. IEEE} \textbf{87}, 606--632 (1999).

\bibitem[{Meschke \emph{et~al.}(2011)Meschke, Engert, Heyer \&
  Pekola}]{Meschke2011}
Meschke, M., Engert, J., Heyer, D. \& Pekola, J.~P.
\newblock Comparison of Coulomb Blockade Thermometers with the International
  Temperature Scale PLTS-2000.
\newblock \emph{Int. J. Thermophys.} \textbf{32}, 1378--1386 (2011).

\bibitem[{Pekola \emph{et~al.}(2013)}]{Pekola2013}
Pekola, J.~P. \emph{et~al.}
\newblock Single-electron current sources: Toward a refined definition of the
  ampere.
\newblock \emph{Rev. Mod. Phys.} \textbf{85}, 1421--1472 (2013).

\bibitem[{Flensberg(1993)}]{Flensberg1993}
Flensberg, K.
\newblock Capacitance and conductance of mesoscopic systems connected by
  quantum point contacts.
\newblock \emph{Phys. Rev. B} \textbf{48}, 11156--11166 (1993).

\bibitem[{Matveev(1995)}]{Matveev1995}
Matveev, K.~A.
\newblock Coulomb blockade at almost perfect transmission.
\newblock \emph{Phys. Rev. B} \textbf{51}, 1743--1751 (1995).

\bibitem[{Nazarov(1999)}]{Nazarov1999}
Nazarov, Y.~V.
\newblock Coulomb Blockade without Tunnel Junctions.
\newblock \emph{Phys. Rev. Lett.} \textbf{82}, 1245--1248 (1999).

\bibitem[{Albrecht \emph{et~al.}(2016)}]{Albrecht2016}
Albrecht, S.~M. \emph{et~al.}
\newblock Exponential protection of zero modes in Majorana islands.
\newblock \emph{Nature} \textbf{531}, 206--209 (2016).

\bibitem[{Kouwenhoven \emph{et~al.}(1991)}]{Kouwenhoven1991}
Kouwenhoven, L. \emph{et~al.}
\newblock Single electron charging effects in semiconductor quantum dots.
\newblock \emph{Z Phys. B} \textbf{85}, 367--373 (1991).

\bibitem[{Staring \emph{et~al.}(1991)}]{Staring1991}
Staring, A. \emph{et~al.}
\newblock Analogies in Optics and Micro-Electronics Coulomb-blockade
  oscillations in a quantum dot.
\newblock \emph{Physica B} \textbf{175}, 226--230 (1991).

\bibitem[{van~der Vaart \emph{et~al.}(1993)}]{vanderVaart1993}
van~der Vaart, N. \emph{et~al.}
\newblock Charging effects in quantum dots at high magnetic fields.
\newblock \emph{Physica B} \textbf{189}, 99--110 (1993).

\bibitem[{Molenkamp \emph{et~al.}(1995)Molenkamp, Flensberg \&
  Kemerink}]{Molenkamp1995}
Molenkamp, L.~W., Flensberg, K. \& Kemerink, M.
\newblock Scaling of the Coulomb Energy Due to Quantum Fluctuations in the
  Charge on a Quantum Dot.
\newblock \emph{Phys. Rev. Lett.} \textbf{75}, 4282--4285 (1995).

\bibitem[{Joyez \emph{et~al.}(1997)Joyez, Bouchiat, Esteve, Urbina \&
  Devoret}]{Joyez1997}
Joyez, P., Bouchiat, V., Esteve, D., Urbina, C. \& Devoret, M.~H.
\newblock Strong Tunneling in the Single-Electron Transistor.
\newblock \emph{Phys. Rev. Lett.} \textbf{79}, 1349--1352 (1997).

\bibitem[{Chouvaev \emph{et~al.}(1999)Chouvaev, Kuzmin, Golubev \&
  Zaikin}]{Chouvaev1999}
Chouvaev, D., Kuzmin, L.~S., Golubev, D.~S. \& Zaikin, A.~D.
\newblock Strong tunneling and Coulomb blockade in a single-electron
  transistor.
\newblock \emph{Phys. Rev. B} \textbf{59}, 10599--10602 (1999).

\bibitem[{Berman \emph{et~al.}(1999)Berman, Zhitenev, Ashoori \&
  Shayegan}]{Berman1999}
Berman, D., Zhitenev, N.~B., Ashoori, R.~C. \& Shayegan, M.
\newblock Observation of Quantum Fluctuations of Charge on a Quantum Dot.
\newblock \emph{Phys. Rev. Lett.} \textbf{82}, 161--164 (1999).

\bibitem[{Duncan \emph{et~al.}(1999)Duncan, Livermore, Westervelt, Maranowski
  \& Gossard}]{Duncan1999}
Duncan, D.~S., Livermore, C., Westervelt, R.~M., Maranowski, K.~D. \& Gossard,
  A.~C.
\newblock Direct measurement of the destruction of charge quantization in a
  single-electron box.
\newblock \emph{Appl. Phys. Lett.} \textbf{74}, 1045--1047 (1999).

\bibitem[{Amasha \emph{et~al.}(2011)}]{Amasha2011}
Amasha, S. \emph{et~al.}
\newblock Coulomb Blockade in an Open Quantum Dot.
\newblock \emph{Phys. Rev. Lett.} \textbf{107}, 216804 (2011).

\bibitem[{Pasquier \emph{et~al.}(1993)}]{Pasquier1993}
Pasquier, C. \emph{et~al.}
\newblock Quantum limitation on Coulomb blockade observed in a 2D electron
  system.
\newblock \emph{Phys. Rev. Lett.} \textbf{70}, 69--72 (1993).

\bibitem[{Crouch \emph{et~al.}(1996)Crouch, Livermore, Westervelt, Campman \&
  Gossard}]{Crouch1996}
Crouch, C.~H., Livermore, C., Westervelt, R.~M., Campman, K.~L. \& Gossard,
  A.~C.
\newblock Coulomb oscillations in partially open quantum dots.
\newblock \emph{Superlattices Microstruct.} \textbf{20}, 377--381 (1996).

\bibitem[{Liang \emph{et~al.}(1998)}]{Liang1998}
Liang, C.-T. \emph{et~al.}
\newblock Experimental Evidence for Coulomb Charging Effects in an Open Quantum
  Dot at Zero Magnetic Field.
\newblock \emph{Phys. Rev. Lett.} \textbf{81}, 3507--3510 (1998).

\bibitem[{Cronenwett \emph{et~al.}(1998)}]{Cronenwett1998}
Cronenwett, S.~M. \emph{et~al.}
\newblock Mesoscopic Coulomb Blockade in One-Channel Quantum Dots.
\newblock \emph{Phys. Rev. Lett.} \textbf{81}, 5904--5907 (1998).

\bibitem[{Tkachenko \emph{et~al.}(2001)}]{Tkachenko2001}
Tkachenko, O.~A. \emph{et~al.}
\newblock Coulomb charging effects in an open quantum dot device.
\newblock \emph{J. Phys Condens Matter} \textbf{13}, 9515 (2001).

\bibitem[{Aleiner \& Glazman(1998)}]{Aleiner1998}
Aleiner, I.~L. \& Glazman, L.~I.
\newblock Mesoscopic charge quantization.
\newblock \emph{Phys. Rev. B} \textbf{57}, 9608--9641 (1998).

\bibitem[{Furusaki \& Matveev(1995)}]{Furusaki1995b}
Furusaki, A. \& Matveev, K.~A.
\newblock Theory of strong inelastic cotunneling.
\newblock \emph{Phys. Rev. B} \textbf{52}, 16676--16695 (1995).

\bibitem[{Yi \& Kane(1996)}]{Yi1996}
Yi, H. \& Kane, C.~L.
\newblock Coulomb blockade in a quantum dot coupled strongly to a lead.
\newblock \emph{Phys. Rev. B} \textbf{53}, 12956--12966 (1996).

\bibitem[{Le~Hur \& Seelig(2002)}]{LeHur2002}
Le~Hur, K. \& Seelig, G.
\newblock Capacitance of a quantum dot from the channel-anisotropic two-channel
  Kondo model.
\newblock \emph{Phys. Rev. B} \textbf{65}, 165338 (2002).

\bibitem[{Matveev \& Andreev(2002)}]{Matveev2002}
Matveev, K.~A. \& Andreev, A.~V.
\newblock Thermopower of a single-electron transistor in the regime of strong
  inelastic cotunneling.
\newblock \emph{Phys. Rev. B} \textbf{66}, 045301 (2002).

\bibitem[{Korshunov(1987)}]{Korshunov1987}
Korshunov, S.~E.
\newblock Coherent and incoherent tunneling in a Josephson junction with a
  `periodic' dissipation.
\newblock \emph{JETP Lett.} \textbf{45}, 434--436 (1987).

\bibitem[{Mitchell \emph{et~al.}(2016)Mitchell, Landau, Fritz \&
  Sela}]{Mitchell2016}
Mitchell, A.~K., Landau, L.~A., Fritz, L. \& Sela, E.
\newblock Universality and Scaling in a Charge Two-Channel Kondo Device.
\newblock \emph{Phys. Rev. Lett.} \textbf{116}, 157202 (2016).


\end{thebibliography}

\normalsize

\begin{thebibliography}{40}
\expandafter\ifx\csname natexlab\endcsname\relax\def\natexlab#1{#1}\fi
\expandafter\ifx\csname url\endcsname\relax
  \def\url#1{\texttt{#1}}\fi
\expandafter\ifx\csname urlprefix\endcsname\relax\def\urlprefix{URL }\fi

\setcounter{NAT@ctr}{30}


\bibitem[{G\"oktas \emph{et~al.}(2008)G\"oktas, Weber, Weis \& von
  Klitzing}]{Goktas2008}
G\"oktas, O., Weber, J., Weis, J. \& von Klitzing, K.
\newblock Alloyed ohmic contacts to two-dimensional electron system in
  AlGaAs/GaAs heterostructures down to submicron length scale.
\newblock \emph{Physica E} \textbf{40}, 1579--1581 (2008).

\bibitem[{Iftikhar \emph{et~al.}(2015)}]{Iftikhar2015}
Iftikhar, Z. \emph{et~al.}
\newblock Two-channel Kondo effect and renormalization flow with macroscopic
  quantum charge states.
\newblock \emph{Nature} \textbf{526}, 233--236 (2015).

\bibitem[{Jezouin \emph{et~al.}(2013)}]{Jezouin2013b}
Jezouin, S. \emph{et~al.}
\newblock Quantum Limit of Heat Flow Across a Single Electronic Channel.
\newblock \emph{Science} \textbf{342}, 601--604 (2013).

\bibitem[{Spietz \emph{et~al.}(2003)Spietz, Lehnert, Siddiqi \&
  Schoelkopf}]{Spietz2003}
Spietz, L., Lehnert, K.~W., Siddiqi, I. \& Schoelkopf, R.~J.
\newblock Primary Electronic Thermometry Using the Shot Noise of a Tunnel
  Junction.
\newblock \emph{Science} \textbf{300}, 1929--1932 (2003).

\bibitem[{Panyukov \& Zaikin(1991)}]{Panyukov1991}
Panyukov, S.~V. \& Zaikin, A.~D.
\newblock Coulomb blockade and nonperturbative ground-state properties of
  ultrasmall tunnel junctions.
\newblock \emph{Phys. Rev. Lett.} \textbf{67}, 3168--3171 (1991).

\bibitem[{Wang \& Grabert(1996)}]{Wang1996}
Wang, X. \& Grabert, H.
\newblock Coulomb charging at large conduction.
\newblock \emph{Phys. Rev. B} \textbf{53}, 12621--12624 (1996).

\bibitem[{Lukyanov(2007)}]{Lukyanov2007}
Lukyanov, S.~L.
\newblock Notes on parafermionic QFTs with boundary interaction.
\newblock \emph{Nucl. Phys. B} \textbf{784}, 151--201 (2007).

\bibitem[{Glazman \& Shekhter(1989)}]{Glazman1989}
Glazman, L.~I. \& Shekhter, R.~I.
\newblock Coulomb oscillations of the conductance in a laterally confined
  heterostructure.
\newblock \emph{J. Phys. Condens. Matter} \textbf{1}, 5811 (1989).

\bibitem[{Idrisov \emph{et~al.}(????)Idrisov, Levkivskyi \&
  Sukhorukov}]{Idrisov2016}
Idrisov, E., Levkivskyi, I. \& Sukhorukov, E. \newblock In preparation.

\bibitem[{Halperin(1982)}]{Halperin1982}
Halperin, B.~I.
\newblock Quantized Hall conductance, current-carrying edge states, and the
  existence of extended states in a two-dimensional disordered potential.
\newblock \emph{Phys. Rev. B} \textbf{25}, 2185--2190 (1982).

\bibitem[{Wen(1990)}]{Wen1990}
Wen, X.~G.
\newblock Chiral Luttinger liquid and the edge excitations in the fractional
  quantum Hall states.
\newblock \emph{Phys. Rev. B} \textbf{41}, 12838--12844 (1990).

\bibitem[{Fr\"ohlich \& Zee(1991)}]{Frohlich1991}
Fr\"ohlich, J. \& Zee, A.
\newblock Large scale physics of the quantum hall fluid.
\newblock \emph{Nucl. Phys. B} \textbf{364}, 517--540 (1991).

\bibitem[{Slobodeniuk \emph{et~al.}(2013)Slobodeniuk, Levkivskyi \&
  Sukhorukov}]{Slobodeniuk2013}
Slobodeniuk, A., Levkivskyi, I. \& Sukhorukov, E.
\newblock Equilibration of quantum Hall edge states by an Ohmic contact.
\newblock \emph{Phys. Rev. B} \textbf{88}, 165307 (2013).

\bibitem[{Sukhorukov(2016)}]{Sukhorukov2016}
Sukhorukov, E.
\newblock Scattering theory approach to bosonization of non-equilibrium
  mesoscopic systems.
\newblock \emph{Physica E: Low-dimensional Systems and Nanostructures}
  \textbf{77}, 191--198 (2016).

\end{thebibliography}
\end{document}